%% file: main_arxiv.tex
\documentclass[a4paper, 12pt]{article} 
\usepackage[a4paper,left=3cm,right=3cm,top=3cm,bottom=3cm]{geometry}
\usepackage{enumerate}  
\usepackage{fancyhdr}
\usepackage{graphicx}
\usepackage{palatino}
\usepackage{multirow}
\usepackage{booktabs}
\usepackage{titlesec}
\usepackage{authblk}
\usepackage{enumitem}
\usepackage[english]{babel}
\usepackage[bookmarks]{hyperref}
\usepackage[justification=centering]{caption}
\usepackage{csquotes}
\usepackage{tikz, pgfplots}
\pgfplotsset{compat=newest}
\usepackage{amsmath,amssymb}
\usepackage{bm}
\usepackage{graphicx,import}
\usepackage{calc}
\usepackage{subcaption, siunitx}
\usepackage{algpseudocode,algorithm,algorithmicx}

\definecolor{vargreen}{rgb}{0.0, 0.5, 0.0}
\definecolor{navyblue}{rgb}{0.0, 0.0, 0.5}
\definecolor{mediumorchid}{rgb}{0.73, 0.33, 0.83}
\definecolor{crimson}{rgb}{0.86, 0.08, 0.24}
\definecolor{lightseagreen}{rgb}{0.13, 0.7, 0.67}
\definecolor{royalblue}{rgb}{0.25, 0.41, 0.88}
\definecolor{hotpink}{rgb}{1.0, 0.41, 0.71}
\definecolor{goldenrod}{rgb}{0.85, 0.65, 0.13}

\begin{document}

\title{A segregated reduced-order model of a pressure-based solver for turbulent compressible flows}
\author[1]{Matteo Zancanaro\footnote{mzancana@sissa.it}}
\author[1]{Valentin Nkana Ngan\footnote{vkanang@sissa.it}}
\author[2]{Giovanni Stabile\footnote{giovanni.stabile@santannapisa.it}}
\author[1]{Gianluigi Rozza\footnote{grozza@sissa.it}}

\affil[1]{Mathematics Area, mathLab, SISSA, via Bonomea 265, I-34136 Trieste, Italy}
\affil[2]{The Biorobotics Institute, Sant'Anna School of Advanced Studies, Viale Rinaldo Piaggio 34, 56025, Pontedera, Pisa, Italy }
\date{\today} 


\maketitle

\begin{abstract}
This article provides a reduced-order modelling framework for turbulent compressible flows discretized by the use of finite volume approaches. The basic idea behind this work is the construction of a reduced-order model capable of providing closely accurate solutions with respect to the high fidelity flow fields. Full-order solutions are often obtained through the use of segregated solvers (\textit{solution variables are solved one after another}), employing slightly modified conservation laws so that they can be decoupled and then solved one at a time. Classical reduction architectures, on the contrary, rely on the Galerkin projection of a complete Navier-Stokes system to be projected all at once, causing a mild discrepancy with the high order solutions. This article relies on segregated reduced-order algorithms for the resolution of turbulent and compressible flows in the context of physical and geometrical parameters. At the full-order level turbulence is modeled using an eddy viscosity approach. Since there is a variety of different turbulence models for the approximation of this supplementary viscosity, one of the aims of this work is to provide a reduced-order model which is independent on this selection. This goal is reached by the application of hybrid methods where Navier-Stokes equations are projected in a standard way while the viscosity field is approximated by the use of data-driven interpolation methods or by the evaluation of a properly trained neural network. 
By exploiting the aforementioned expedients it is possible to predict accurate solutions  with respect to the full-order problems characterized by high Reynolds numbers and elevated Mach numbers.

\end{abstract}


\input{sections/intro}

\input{sections/compressibleNS}

\input{sections/FVdiscretization}
\input{sections/pressureEq}
\section{Reduced-order modeling architecture}
\input{sections/POD}
\input{sections/compRedSimple}
\input{sections/turbTreatment}
\section{Numerical results}
\input{sections/physical}

\input{sections/geometrical}
\input{sections/conclusions}
\section*{Data availability}
The data sets generated during and/or analyzed during the current study are
available from the corresponding authors on reasonable request.
\section*{Disclosure statement}
The authors report no potential conflict of interest.
\input{sections/aknw}

\newpage

\addcontentsline{toc}{chapter}{\bibname}
\bibliographystyle{ieeetr}
\bibliography{bib/references}
\end{document}

%% file: sections/intro.tex
\section{Introduction}
In the last decades fluid flow simulations have progressively enlarged their applicability and their influence in many different research fields (general overviews can be found in \cite{anderson1995computational,wendt2008computational,blazek2015computational}. Nowadays, Computational Fluid Dynamics (CFD) applications have reached widely spread ambits as, for example, shape optimization for naval/automotive/aerospace engineering \cite{jameson2003aerodynamic,samareh2004aerodynamic}, cardiovascular in real time surgery \cite{formaggia2010cardiovascular}, chemistry industrial processes \cite{murthy2007cfd,van2004cfd} or weather forecasts \cite{castellani2014wind}. While increasing the demand on reliability and usability of CFD methods, the computational capability of the employed hardware architectures are no more sufficient in terms of time consuming. For this reason, the search for new efficient methods able to reduce computational time keeps on covering a relevant amount of CFD research efforts.

A popular research field, related to the aforementioned applications, is the analysis of the dynamics relative to parameterized Partial Differential Equations (PDEs). In this case an infinite number of solutions is available for every slightly different value of the parameter. For some specific ambits, many of them have to be evaluated in order to find out the one that is best performing in terms of prefixed requirements fulfilling. The cost of such a procedure can easily become unaffordable, in particular when the dimension of the problem is big enough \cite{morozova2020feasibility}. Typical applications where such a difficulty is often arising are, for example, shape optimization problems, uncertainty quantification studies or optimal control frameworks.

Recently many different techniques have been taken into consideration to try to overtake this problem. Galerkin projection has widely been employed to develop new reduction strategies capable of exploiting the information of just a few full order solutions for different parameter values in order to perform efficient, accurate and much cheaper solutions for a different selection of the parameter (for fluid flow applications, \cite{amsallem2008interpolation,tezaur2017advanced,yano20206,weller2010numerical,StabileRozza2018} represent relevant works). Many different choices are possible to take advantage of the dynamical content contained in high fidelity solutions. The most used ones are the Proper Orthogonal Decomposition (POD) \cite{akhtar2009stability,baiges2014reduced,burkardt2006pod,kunisch2002galerkin,san2013proper}, the Proper Generalized Decomposition (PGD) \cite{dumon2013proper,chinesta2011short}, the Dynamic Mode Decomposition (DMD) \cite{schmid2010dynamic,kutz2016dynamic} or the Greedy algorithm \cite{hesthaven2016certified,urban2014greedy}. The first idea behind the POD, as it was originally formulated in the domain of fluid dynamics to analyze turbulence, is to decompose a vector field into a set of deterministic spatial functions weighted by time/parameters coefficients.  

Recently, Machine Learning (ML) has emerged as a new branch of research in this field. By the use of neural networks (NNs),  the resolution of complex non-linear parametric PDEs has become easier and more accessible. In the following studies \cite{hesthaven2018non,wang2019non,mohan2018deep,mannarino2014nonlinear,PapapiccoDemoGirfoglioStabileRozza2021,RomorStabileRozza2022}, this assertion has been tested by combining POD and NNs method to a wide range of applications. For instance, to the non-linear Poisson equation in one and two spatial dimensions, and on two-dimensional cavity viscous flows, modeled through the steady incompressible Navier-Stokes equations.
Both the two aforementioned approaches have some valuable aspects together with shortcomings to be underlined. Projection techniques are strongly connected with physical laws of the problem since they use modal basis functions obtained by real solutions to extract the main dynamics and they employ those modes to project and reconstruct conservation equations solutions manifolds. Unfortunately non-linearity and non-affinity of the parameterized formulation can be difficult issues to be carefully treated. Moreover,  sometimes,  it may happen that the equations are not directly available and in that case these methods are not employable. A classical example is constituted by commercial software where a deep description of the employed laws is not provided. 
Conversely to projection techniques, ML techniques are very versatile. They only require a set of trained solutions despite the complexity of the mathematical formulation of the  problem at hand. Those techniques are constructed and modeled to yield good approximations in a short time.
The dark side of these approaches is the fact that they have a much weaker connection with the real physics of what they are approximating and the actual meaning of every single part constituting their architecture is arguably comprehensible in terms of phenomena representation. For this reason they may give inaccurate results thanks to impossibility in having a deeper check on networks responses. 

Taking all the aforementioned examinations under consideration, this work provides a new mixed technique for compressible Navier-Stokes  problems, capable of merging the advantages of projection techniques together with data-driven architectures. In particular, in our approach, classical projection methods are used for the Favre Averaged Navier Stokes (FANS) equations while a neural network gets trained to provide the eddy viscosity solutions in a turbulence modeling approach.  These new contributions result to a reduced-order models that are independent of the selection of turbulence models for any segregated solvers for compressible flows capable to reduce the computational cost associated with fluid flow problems characterized by high Reynolds numbers and elevated Mach numbers.
The main goal is to propose an architecture proficient in dealing with different types of parametrizations for compressible flows. Moreover one of the most relevant focuses concerning this work is constituted by a coherent approach between full-order and reduced-order solutions, by developing a new reduced compressible SIMPLE (Semi-Implicit Method for Pressure Linked Equations) algorithm. 

This manuscript is structured in  six different sections. The \autoref{sec:compNS}, and  \autoref{sec:fvDisc} present the equations  used in  this work and their Finite Volume Method (FVM); \autoref{sec:pod} explains the POD procedure employed to obtain the modal basis functions. In \autoref{sec:redComp} the core algorithm used for our technique is introduced together with \autoref{sec:turbTr} where the AI architecture for turbulence treatment is shown. Two different test cases, a physically parameterized and a geometrically parameterized ones, are exposed in \autoref{sec:phy} and \autoref{sec:geom} respectively. Finally,  in \autoref{sec:conclusions},  few considerations on the results and some possible developments for this work are presented.

%% file: sections/compressibleNS.tex
\section{The compressible Navier-Stokes equations}\label{sec:compNS}

In this work we want to deal with parameterized compressible Navier-Stokes equations problems. To manage the compressibility of the fluid, we selected a common strategy for this kind of applications: the Favre averaging. The equations describing the physics are the following ones:
\begin{align} \label{eq:COMP}
\begin{cases}
    \displaystyle \frac{\partial \rho}{\partial t} + \nabla \cdot \left(\rho \bm{u}\right)= 0 ~~ \text{in} \ \Omega(\pi),\\
    \displaystyle \frac{\partial \rho \bm{u} }{\partial t} + \nabla \cdot \left[ \rho \bm{u} \otimes \bm{u}+ p \bm{I} - \bm{\tau} \right] = 0 ~~ \text{in} \ \Omega(\pi), \\
    \displaystyle \frac{\partial \rho e_0}{\partial t} + \nabla \cdot \left[ \rho \bm{u}e_0 + p\bm{u} - \bm{u} \cdot\bm{\tau} - q \right] = 0 ~~ \text{in} \ \Omega(\pi),\\
    \bm{u} = \bm{g}_D ~~ \text{in} \ \Gamma_D,\\
    \nu \displaystyle \frac{\partial \bm{u}}{\partial \bm{n}}-p \bm{n} = g_N ~~\text{in} \ \Gamma_{N},
\end{cases}    
\end{align}
where $\rho$ indicates the density, $\bm{u}$ the flow velocity, $p$ the pressure,  $\bm{\tau}$  the viscous stress tensor, $e_0$ the total energy, and $\bm{I}$ the identity tensor. $\Gamma_D$ stands for the part of the boundary where the Dirichlet condition $\bm{g}_D$ is imposed while $\Gamma_N$ is the part of the boundary where the Neumann condition $\bm{g}_N$ is imposed, $\nu$ the kinematic viscosity, $\bm{n}$ the unit normal vector, and $\Omega(\pi)$ is the computational domain and it can be, in geometrical parametrization cases directly dependent on the parameter $\pi$.
The heat-flux $q$ is given by Fourier's law: 
\begin{eqnarray}
q = - \lambda \nabla T \equiv C_p\frac{\mu}{Pr}\nabla T; 
\end{eqnarray}
the laminar  Prandtl number $Pr$ is given by: $Pr = \frac{C_p\mu}{\lambda}$.
To close these equations it is also necessary to specify an equation of state. Assuming air to be an ideal gas,  the following relations are valid: $$\gamma \equiv C_p/C_v, ~~, p = \rho R T, ~~, e = C_vT, ~~, C_p - C_v = R. $$
Being $R$ the gas constant, $C_v$ is the constant volume, and $C_p$ means specific heat at constant pressure, $\gamma$ is the adiabatic index, $e$ the internal energy, and $T$ the temperature.
In the Favre Averaged Navier-Stokes (FANS) equations, all the variables (density $\rho$, pressure $p$, velocity $\bm{u}$, total energy $e_0$, temperature $T$ and internal energy $e$) are decomposed in an averaged part and a fluctuating one as follows:
\begin{eqnarray}
\rho = \overline{\rho} + \rho', ~~~ p = \overline{p} + p', ~~~  T = \tilde{T} + T'' \label{favre1} \\ 
e_0 = \tilde{e_0} + e_0'', ~~~ \bm{u} = \tilde{\bm{u}} + \bm{u}'', ~~~ e = \tilde{e} + e'' \label{favre2} .
\end{eqnarray}
Superscript $\tilde{\square}$ indicates the Favre averaging which correspond to a density weighted Reynolds averaging $\overline{\square}$. Given a certain variable $\Phi(t)$, we have:
\begin{align}
\overline{\Phi}  & = \frac{1}{T} \int_T \Phi(t) dt \Rightarrow \Phi' = \Phi - \overline{\Phi} \label{fv1} \\ 
\Tilde{\Phi}   & = \frac{\overline{\rho \Phi}}{\overline{\rho}} \Rightarrow \Phi'' = \Phi - \tilde{\Phi} \label{fv2}.
\end{align}
Plugging \autoref{favre1}, \autoref{favre2}, \autoref{fv1} and \autoref{fv2} in \autoref{eq:COMP} lead to:
\begin{align} \label{eq:FANS}
\begin{cases}
    \displaystyle \frac{\partial \overline{\rho}}{\partial t} + \nabla \cdot \left( \overline{\rho} \tilde{\bm{u}}\right)= 0 ~~ \text{in} \ \Omega(\pi),\\
    \displaystyle \frac{\partial \overline{\rho} \tilde{\bm{u}}}{\partial t} + \nabla \cdot \left[ \overline{\rho} \tilde{\bm{u}} \otimes \tilde{\bm{u}} - \tilde{\bm{\tau}}_{turb} - \tilde{\bm{\tau}} + \overline{p} \bm{I}\right] = 0 ~~ \text{in} \ \Omega(\pi),\\
    \displaystyle \frac{\partial \overline{\rho} \tilde{e}_0}{\partial t} + \nabla \cdot \left[ \overline{\rho} \tilde{\bm{u}} \tilde{e}_0 - C_p \left(\frac{\mu}{Pr} + \frac{\mu_t}{Pr_{t}}\right)\nabla \tilde{T} \right] 
    \\ + \nabla \cdot \left[\overline{p} \tilde{\bm{u}} - \tilde{\bm{u}} \cdot \tilde{\bm{\tau}}  - \tilde{\bm{u}} \cdot \tilde{\bm{\tau}}_{turb} \right] = 0 ~~ \text{in} \ \Omega(\pi),\\
    \tilde{\bm{u}} = \bm{g}_D ~~ \text{in} \ \Gamma_D,\\
    \nu \displaystyle \frac{\partial \tilde{\bm{u}}}{\partial \bm{n}} - \overline{p} \bm{n} = g_N ~~\text{in} \ \Gamma_{N},
\end{cases}    
\end{align}
where $\overline{p}$, $\tilde{\bm{u}}$ and $ \tilde{e}$ become the unknowns of the problem.
$\bm{\tau}_{turb}$ stands for the extra viscosity term due to turbulence, $\mu$ is the dynamic viscosity, $\mu_t$ is the extra viscosity owing to turbulence, $Pr$ indicates the Prandtl number and $Pr_t$ its turbulent counterpart which is a constant value. The molecular $\tilde{\bm{\tau}} $ and Reynolds-Stress $\tilde{\bm{\tau}}_{turb}$ tensors are given by:
\begin{eqnarray}
\tilde{\bm{\tau}} = 2\mu \tilde{\bm{S}}, \hspace{0.05cm} \tilde{\bm{\tau}}_{turb} = 2\mu_t \tilde{\bm{S}} -\frac{2}{3}\Bar{\rho}k\bm{I},
\end{eqnarray}
where $\tilde{\bm{S}} = \frac{\nabla \tilde{\bm{u}} + \nabla \tilde{\bm{u}}^T}{2} - \frac{1}{3} \nabla \cdot \tilde{\bm{u}} \bm{I}$, and 
$k = \widetilde{\frac{\bm{u}'' \cdot \bm{u}''}{2}}$. Moreover, the density averaged total energy $\Tilde{e}_0$  is rewritten in the internal energy form:
\begin{eqnarray}
\Tilde{e}_0    = \Tilde{e} + \frac{\tilde{\bm{u}} \cdot \tilde{\bm{u}}}{2} + k,
\end{eqnarray}

\autoref{eq:FANS} is obtained after some approximations and assumptions from an eddy viscosity point of view. The reader interested in the averaging procedure and modeling should refer to \cite{wilcox1998turbulence}.

From now on, \autoref{eq:FANS} will be considered only in its steady-state formulation.
All the averaged variables are dependent on the parameter $\pi$ but,  for the  sake of simplicity, the following notation will be used: 
$$\overline{\rho} = \overline{\rho}(\pi), \hspace{0.1cm} \overline{p} = \overline{p}(\pi), \hspace{0.1cm} \tilde{\bm{u}} = \tilde{\bm{u}}(\pi), \hspace{0.1cm} \tilde{T} = \tilde{T}(\pi), \hspace{0.1cm} \tilde{e} = \tilde{e}(\pi).$$

In the energy equation, the viscous terms are neglected in many solvers.  This, can be reasonably true if compared with the other terms present into the energy equation. Moreover, the turbulent kinetic energy is neglected in the total energy. 
This  results in the following system:

\begin{equation} \label{eq:rhosimple}
    \begin{cases}
        \nabla \cdot \left( \overline{\rho} \tilde{\bm{u}}\right)= 0 ~~~ \text{in} \ \Omega(\pi),\\
        \displaystyle \nabla \cdot \left[ \overline{\rho} \tilde{\bm{u}} \otimes \tilde{\bm{u}} - \mu_{eff} \bigg( \nabla \tilde{\bm{u}} + \nabla \tilde{\bm{u}}^T
        - \frac{2}{3} \nabla \cdot \tilde{\bm{u}} \bm{I}\bigg) + \overline{p} \bm{I}\right]
        = 0 ~~~ \text{in} \ \Omega(\pi),\\
        \displaystyle \nabla \cdot \bigg[ \overline{\rho} \tilde{\bm{u}} \bigg( \tilde{e} + \frac{\tilde{\bm{u}} \cdot \tilde{\bm{u}}}{2}\bigg) - \alpha_{eff} \nabla \tilde{e} + \overline{p} \tilde{\bm{u}}  \bigg]  = 0 ~~~ \text{in} \ \Omega(\pi),\\
        \tilde{\bm{u}} = \bm{g}_D ~~~ \text{in} \ \Gamma_D,\\
        \nu \displaystyle \frac{\partial \tilde{\bm{u}}}{\partial \bm{n}} - \overline{p} \bm{n} = g_N  ~~~ \text{in} \ \Gamma_{N}.
    \end{cases}    
\end{equation}
With $\mu_{eff} = \mu + \mu_t$, and $\alpha_{eff} = \gamma \bigg( \frac{\mu}{Pr} + \frac{\mu_t}{Pr}_{t} \bigg)$. It is now clear that all the turbulence-related terms of the equations rely on $\mu_t$ to be calculated. For this reason, since only the eddy viscosity is required, a common 2-equations turbulent model as, e.g., $k-\epsilon$ or $k-\omega$ \cite{wilcox1998turbulence}, is sufficient as a closure for the problem.

%% file: sections/FVdiscretization.tex
\section{Full-order discretization method}\label{sec:fvDisc}
The first step towards a Finite Volume discretization (for a deeper insight see \cite{eymard2000finite}) of the problem is the division of the domain $\Omega(\pi)$ into a tessellation $\mathcal{T}(\pi)$ composed by a certain number $N_h$ of cells $\Omega_i(\pi)$, so that:
\begin{equation*}
    \mathcal{T}(\pi) = \{ \Omega_i(\pi) \}_{i=1}^{N_h}, \hspace{1cm} \bigcup_{i=1}^{N_h} \Omega_i (\pi) = \Omega (\pi),
\end{equation*}
where every cell $\Omega_i$ can be constructed as a non-convex polyhedron.

The Finite Volume variables can be here introduced: $\overline{p}_h \in \mathbb{Q}_h$, $\tilde{\bm{u}}_h \in \mathbb{V}_h$ and $\tilde{e}_h \in \mathbb{E}_h$. They are not continuous and they are constant in the interior part of each cell assuming everywhere the value at the center of the cell. For sake of simplicity in this section we will keep on referring to this variables without the $\square_h$ subscript to not make the formulas too heavy.
\subsection{Finite Volume discretization}\label{sec:fvm}
\begin{figure}
\centering
   \includegraphics[width=0.65 \textwidth]{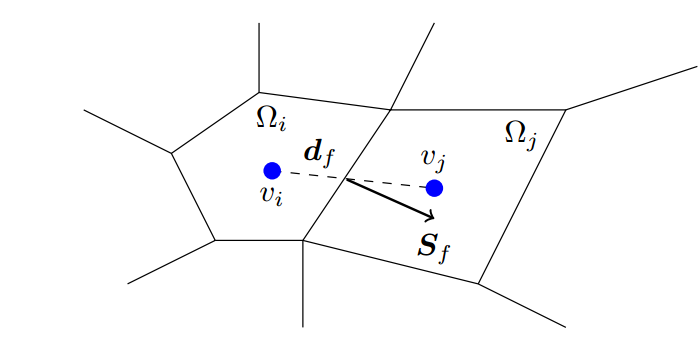}
    \caption{Relation between two neighbor cells of the tessellation $\mathcal{T}$ for a certain variable $v$.}
    \label{fig:cells}
\end{figure}
This work uses a segregated approach based on a compressible formulation of the SIMPLE algorithm.  This means that the equations for each variable characterizing the system (the velocity, the pressure, the energy (either internal energy or enthalpy depending on the choice of the user) and the variables characterizing turbulence) are solved sequentially and the solution of the previous equations is inserted into the subsequent equation. This aspect has to be kept in mind for the Finite Volume discretization strategy.
\textit{A main advantage of a  segregated algorithm is the  memory-efficiency, since the discretized equations need only to be stored in memory one at a time. However, a shortcoming of the segregated approach is the slowly convergence of the solution as the equations are solved in a decoupled manner.}
We can now take into consideration the equations one by one, starting from the continuity constrain. For a detailed treatment of what concerns fluid flows equations discretization, the interested reader can refer to \cite{moukalled2016finite}. The continuity equation can be written in its integral form over each cell as follows:
\begin{equation}
    \int_{\Omega_i} \nabla \cdot \left( \overline{\rho} \tilde{\bm{u}} \right) dV = 0.
\end{equation}
By exploiting the divergence theorem, the discretized version of the continuity equation then reads:
\begin{equation}
    \int_{\delta \Omega_i} \overline{\rho} \tilde{\bm{u}} \cdot d\bm{S} \simeq \sum_{f=1}^{Nf_i} \left( \overline{\rho} \tilde{\bm{u}} \right)\rvert_{f} \cdot \bm{S}_f =  \sum_{f=1}^{Nf_i} F_f = 0,
\end{equation}
where the subscript $\square\rvert_{f}$ indicates that those variables are evaluated at the center of the face $f$ and $\bm{S}_f$ is the oriented surface of the same face while $Nf_{i}$ is the total number of faces surrounding the $i$-th cell while $F_f$ stands for the mass flux crossing the face $f$ as shown in \autoref{fig:cells}. 
It is important to  recall that, all the variables in a Finite Volume scheme are known only at the cell center,  so their values at the center of the faces has to be calculated by interpolating the neighbor cell center values.

Let us now take into consideration the momentum equation. It has to be integrated over the volume of every cell and it can then be analysed term by term, starting from the convective one:
\begin{align}
    \int_{\Omega_i} \bm{\nabla} \cdot (\overline{\rho}\tilde{\bm{u}} \otimes \tilde{\bm{u}}) \, dV  
    & \simeq 
    \sum_{f=1}^{Nf_i} \bm{S}_f \cdot \left( \overline{\rho} \tilde{\bm{u}} \otimes \tilde{\bm{u}}\right)\rvert_f = \sum_{f=1}^{{Nf}_i} \bm{F}_f \tilde{\bm{u}}_f.
\end{align}
The first part of the diffusion term is discretized as follows:
\begin{align}
    \int_{\Omega_i} \bm{\nabla} \cdot \left[\mu_{eff} \nabla \tilde{\bm{u}} \right]dV 
    & \simeq \displaystyle\sum_{f=1}^{Nf_i} \left[ \mu_{eff} \nabla \tilde{\bm{u}} \right] \rvert_f \cdot \bm{S}_f.
\end{align}
For orthogonal meshes we can approximate this term as:
\begin{equation}
\sum_{f=1}^{Nf_i} \left[\mu_{eff} \nabla \tilde{\bm{u}} \right] \rvert_f \cdot \bm{S}_f \simeq \mu_{eff} \rvert_f \ \lvert \bm{S}_f \rvert \frac{\tilde{\bm{u}}_i - \tilde{\bm{u}}_j}{\lvert \bm{d}_f \rvert},
\end{equation}
being $\bm{d}_f$ the oriented vector bridging the cell centers of two neighbor cells. When this is  not the case, a non-orthogonal correction is added:
\begin{align}
\sum_{f=1}^{Nf_i} \left[\mu_{eff} \nabla \tilde{\bm{u}} \right] \rvert_f \cdot \bm{S}_f & \simeq
\sum_{f=1}^{Nf_i} \mu_{eff} \rvert_f\left[ \lvert \bm{\mathcal{P}}_f \rvert \frac{\tilde{\bm{u}}_i - \tilde{\bm{u}}_j}{\lvert \bm{d} \rvert} + \bm{\mathcal{O}}_f \cdot \nabla \tilde{\bm{u}}\rvert_f \right],
\end{align}
where $\bm{\mathcal{P}}_f \parallel \bm{d}_f$, $\bm{\mathcal{O}}_f \bot \bm{d}_f$ and $\bm{\mathcal{P}}_f + \bm{\mathcal{O}}_f = \bm{S}_f$ while $\nabla \tilde{\bm{u}}\rvert_f $ is evaluated starting from its value at the cell centers $\nabla \tilde{\bm{u}}_i$ and $\nabla \tilde{\bm{u}}_j$ by interpolation.

The second part of the diffusion term is treated following the previous steps:
\begin{align}
    \int_{\Omega_i} \bm{\nabla} \cdot \left[\mu_{eff} \nabla \tilde{\bm{u}}^T \right]dV  
    & \simeq 
    \sum_{f=1}^{Nf_i} \left[ \mu_{eff} \nabla \tilde{\bm{u}}^T \right] \rvert_f \cdot \bm{S}_f.
\end{align}
In this case, the face center evaluation is treated explicitly so that this term is considered to be a forcing term:
\begin{align}
    \nabla \tilde{\bm{u}}^T \rvert_f \cdot \bm{S}_f  & = 
    \begin{bmatrix}
    \frac{\partial \tilde{u}_x}{\partial x} S_x + \frac{\partial \tilde{u}_y}{\partial x} S_y + \frac{\partial \tilde{u}_z}{\partial x} S_z\\
    \frac{\partial \tilde{u}_x}{\partial y} S_x + \frac{\partial \tilde{u}_y}{\partial y} S_y + \frac{\partial \tilde{u}_z}{\partial y} S_z\\
    \frac{\partial \tilde{u}_x}{\partial z} S_x + \frac{\partial \tilde{u}_y}{\partial z} S_y + \frac{\partial \tilde{u}_z}{\partial z} S_z\\
    \end{bmatrix}.
\end{align}
The same applies for the last part of the diffusive term:
\begin{align}
    \int_{\Omega_i} \bm{\nabla} \left[\mu_{eff} \frac{2}{3} \nabla \cdot \tilde{\bm{u}} \right] dV 
    & \simeq 
    \sum_{f=1}^{Nf_i} \left[ \mu_{eff} \frac{2}{3} \nabla \cdot \tilde{\bm{u}} \right] \Big\rvert_f \bm{S}_f,
\end{align}
where once again the divergence of the velocity is interpolated to the surface and treated explicitly leading to an additional forcing term.

The last term to be considered is the pressure gradient:
\begin{equation}
    \int_{\Omega_i} \bm{\nabla} \overline{p} \, dV = \int_{\delta \Omega_i} \overline{p} \ d\bm{S} \simeq \sum_{f=1}^{Nf_i} \overline{p}_f \bm{S}_f,
\end{equation}
In the momentum equation, pressure is interpolated to the faces and then treated explicitly as a source term. The final momentum equation reads:
\begin{align}\label{eq:momEq}
    & \sum_{f=1}^{Nf_i} \left[ \bm{F}_f \tilde{\bm{u}}_f - \mu_{eff} \rvert_f \left( \lvert \bm{\mathcal{P}}_f \rvert \frac{\tilde{\bm{u}}_i - \tilde{\bm{u}}_j}{\lvert \bm{d} \rvert} + \bm{\mathcal{O}}_f \cdot \nabla \tilde{\bm{u}}\rvert_f \right) \right] \nonumber \\ 
    & = \sum_{f=1}^{Nf_i} \left[\mu_{eff}\rvert_f \left( \nabla \tilde{\bm{u}}^T_f \cdot \bm{S}_f - \frac{2}{3} \nabla \cdot \tilde{\bm{u}}_f \bm{S}_f\right) - \overline{p}_f \bm{S}_f \right]
\end{align}
where all the terms composing the right-hand side of the equation are treated explicitly as source terms.
\autoref{eq:momEq} can be rewritten in its Finite Volume matrix form as follows:
\begin{equation}\label{eq:mMomEq}
    \bm{A}_u(\tilde{\bm{u}}) = -\nabla \overline{p} \Rightarrow \bm{A} \tilde{\bm{u}} = \bm{H} (\tilde{\bm{u}}) -\nabla \overline{p},
\end{equation}
where $\bm{A}_u(\tilde{\bm{u}})$ is the Finite Volume discretized form containing all the terms related to velocity of both left-hand and right-hand sides of \autoref{eq:momEq}, $\bm{A}\tilde{\bm{u}}$ is the diagonal part of $\bm{A}_u(\tilde{\bm{u}})$ while $-\bm{H} (\overline{\bm{u}})$ is its extra diagonal part so that $\bm{A}_u(\tilde{\bm{u}}) = \bm{A}\tilde{\bm{u}} - \bm{H} (\tilde{\bm{u}})$.

The last equation to be analysed regards the energy conservation:
\begin{equation}
    \int_{\Omega_i} \nabla \cdot \left[ \overline{\rho} \tilde{\bm{u}} \tilde{e} \right] \, dV = \int_{\delta \Omega_i} \overline{\rho} \tilde{\bm{u}} \tilde{e} \cdot d\bm{S} \simeq \sum_{f=1}^{Nf_i} \tilde{e}_f \overline{\rho}_f \tilde{\bm{u}}_f \cdot \bm{S}_f = \sum_{f=1}^{Nf_i} \tilde{e}_f \overline{\rho}_f F_f.
\end{equation}
The kinetic part of the total energy is treated explicitly and leads to:
\begin{align}
    \int_{\Omega_i} \nabla \cdot \left[ \overline{\rho} \tilde{\bm{u}} \frac{\tilde{\bm{u}} \cdot \tilde{\bm{u}}}{2} \right] dV 
    & \simeq 
    \sum_{f=1}^{Nf_i} \frac{\tilde{\bm{u}}_f \cdot \tilde{\bm{u}}_f}{2} \overline{\rho}_f \tilde{\bm{u}}_f \cdot \bm{S}_f
\end{align}
The diffusive term reads:
\begin{eqnarray}
\int_{\Omega_i} \nabla \cdot \left[ \alpha_{eff} \nabla \tilde{e} \right]dV 
\simeq 
\sum_{f=1}^{Nf_i} \alpha_{eff}\Big \rvert_f \nabla \tilde{e}_f \cdot \bm{S}_f.
\end{eqnarray}
Once again the energy gradient is not available at the center of the faces but it can be approximated:
\begin{align}
    & \sum_{f=1}^{Nf_i} \alpha_{eff}\Big \rvert_f \nabla \tilde{e}_f \cdot \bm{S}_f\\
    & \simeq \sum_{f=1}^{Nf_i} \alpha_{eff}\Big \rvert_f \left[ \lvert \bm{\mathcal{P}}_f \rvert \frac{\tilde{e}_i - \tilde{e}_j}{\lvert \bm{d} \rvert} + \bm{\mathcal{O}}_f \cdot \nabla \tilde{e}\rvert_f \right].
\end{align}
Finally, the pressure term is discretized and treated explicitly:
\begin{equation}
    \int_{\Omega_i} \nabla \cdot \left[ \overline{p} \tilde{\bm{u}} \right] \, dV = \int_{\delta \Omega_i} \overline{p} \tilde{\bm{u}} \cdot d\bm{S} \simeq \sum_{f=1}^{Nf_i} \overline{p}_f \tilde{\bm{u}}_f \cdot \bm{S}_f = \sum_{f=1}^{Nf_i} \frac{\overline{p}_f}{\overline{\rho}_f} F_f.
\end{equation}
The resulting equation reads:
\begin{align}\label{eq:enEq}
    & \sum_{f=1}^{Nf_i} \bigg[ \tilde{e}_f F_f - \alpha_{eff}\Big \rvert_f * \bigg( \lvert \bm{\mathcal{P}}_f \rvert \frac{\tilde{e}_i - \tilde{e}_j}{\lvert \bm{d} \rvert} + \bm{\mathcal{O}}_f \cdot \nabla \tilde{e}\rvert_f \bigg) \bigg] \nonumber \\
    & = - \sum_{f=1}^{Nf_i} \bigg(  \frac{\tilde{\bm{u}}_f \cdot \tilde{\bm{u}}_f}{2} + \frac{\overline{p}_f}{\overline{\rho}_f} \bigg) F_f. 
\end{align}
Also \autoref{eq:enEq} can be written into its matrix form as follows:
\begin{equation}\label{eq:mEnEq}
    \bm{E}(\tilde{e}) = \bm{F}(\overline{p},\tilde{\bm{u}}).
\end{equation}

%% file: sections/pressureEq.tex
\subsection{Pressure equation for compressible flows}\label{sec:pressureEq}

By following what has been done in \cite{jasak1996error}, let us localize \autoref{eq:mMomEq} at a generic $\Omega_i$ cell center, we get:
\begin{equation}
    \tilde{\bm{u}}_i = \frac{\bm{H}(\tilde{\bm{u}})}{a_i}  - \frac{\nabla \overline{p}_i}{a_i}.
\end{equation}
Let us rename $\tilde{\bm{u}} = \tilde{\bm{u}}^*$ and $\overline{\rho} = \overline{\rho}^*$ both velocity and density we have at this point, after having solved the momentum equation, for a reason that will be clarified in a moment.
The mass flux, at the generic cell center, can be obtained as:
\begin{equation*}
    \overline{\rho}_i^*\tilde{\bm{u}}_i^* = \overline{\rho}_i^* \frac{\bm{H}(\tilde{\bm{u}}^*)}{a_i}  - \overline{\rho}_i^* \frac{\nabla \overline{p}^{n-1}_i}{a_i}.
\end{equation*}
Since the pressure gradient has to be calculated explicitly, we indicate it as $\nabla \overline{p}^{n-1}$ meaning that the pressure field has to be previously calculated.

It is easy to realize that the set $\overline{\rho}^*, \tilde{\bm{u}}^*, \overline{p}^{n-1}$ will not satisfy the mass conservation constrain since velocity field has been evaluated by the use of the pressure gradient at step $n-1$. We can then imagine to introduce some corrections to all the terms so that $\overline{\rho} = \overline{\rho}^* + \overline{\rho}', \tilde{\bm{u}} = \tilde{\bm{u}}^* + \tilde{\bm{u}}', \overline{p} = \overline{p}^{n-1} + \overline{p}'$.

It is now possible to rewrite the mass flux as:
\begin{align*}
    & \left(\overline{\rho}_i^* + \overline{\rho}_i'\right)\left(\tilde{\bm{u}}_i^* + \tilde{\bm{u}}_i'\right) \\
    & = \left(\overline{\rho}_i^* + \overline{\rho}_i'\right) \left[\frac{\bm{H}(\tilde{\bm{u}}^*)}{a_i} + \frac{\bm{H}(\tilde{\bm{u}}')}{a_i} \right] - \left(\overline{\rho}_i^* + \overline{\rho}_i'\right) \left[\frac{\nabla \overline{p}^{n-1}_i}{a_i} + \frac{\nabla \overline{p}_i'}{a_i} \right].
\end{align*}
By the definition of compressibility $\Psi$, we can write 
$\overline{\rho} = \Psi \overline{p}$ and then $\rho = \overline{\rho}^* + \overline{\rho}' = \Psi \overline{p}^{n-1} + \Psi \overline{p}'.$

Thus, $\overline{\rho}' = \Psi p - \Psi \overline{p}^{n-1} = \Psi \overline{p}'.$
We can then interpolate that expression to obtain the variables evaluations at the faces and finally sum over all the faces surrounding the cell $\Omega_i$ to get the mass conservation equation in its pressure correction shape:
\begin{align*}
& \sum_{f=1}^{Nf_i}\left(\overline{\rho}_i^* + \Psi \overline{p}_i' \right) \left[\frac{\bm{H}(\tilde{\bm{u}}^*)}{a_i} + \frac{\bm{H}(\tilde{\bm{u}}')}{a_i} \right] \Bigg\rvert_f  -\\
& \sum_{f=1}^{Nf_i}\left(\overline{\rho}_i^* + \underbrace{\overline{\rho}_i'}_{\star}\right) *\left[\frac{\nabla \overline{p}^{n-1}_i}{a_i} + \frac{\nabla \overline{p}_i'}{a_i} \right]\Bigg\rvert_f = 0.
\end{align*}
The $\star$ term can be neglected obtaining the correction equation for pressure. The only term that has to be modeled in some way is $\bm{H}(\tilde{\bm{u}}')$. In the SIMPLE-based algorithms the correction extra diagonal velocity term is neglected leading to the following final pressure correction equation:
\begin{equation}\label{eq:mPresEq}
\sum_{f=1}^{Nf_i}\left(\overline{\rho}_i^* + \Psi \overline{p}_i' \right) \left[\frac{\bm{H}(\tilde{\bm{u}}^*)}{a_i} \right] \Bigg\rvert_f  = \sum_{f=1}^{Nf_i}\left(\overline{\rho}_i^* \right) \left[\frac{\nabla \overline{p}^{n-1}_i}{a_i} + \frac{\nabla \overline{p}_i'}{a_i} \right]\Bigg\rvert_f.
\end{equation}

%% file: sections/POD.tex

\subsection{Proper Orthogonal Decomposition procedure}\label{sec:pod}
The scope of this work is to find an efficient and reliable reduced order model to be able to solve \autoref{eq:FANS} for many different values of the parameter $\pi$ without solving the Finite Volume discretized equations every time from scratch. For this reason,  we developed a new procedure based on a POD-Galerkin scheme. 

The whole machinery is divided in two main steps: an offline phase which consists on the resolution of a certain number $N_{\pi}$ of full-order solutions, trying to extract as much information as possible from this set, and an online phase consisting on the resolution of a dimensionally reduced problem for all the different needed parametric configurations. What is new in this method is to be capable of resulting as general as possible with respect to the selected full-order turbulence model and, at the same time, as coherent as possible with respect to high fidelity solutions.

\begin{figure}
    \centering
    \includegraphics[width=0.5\textwidth]{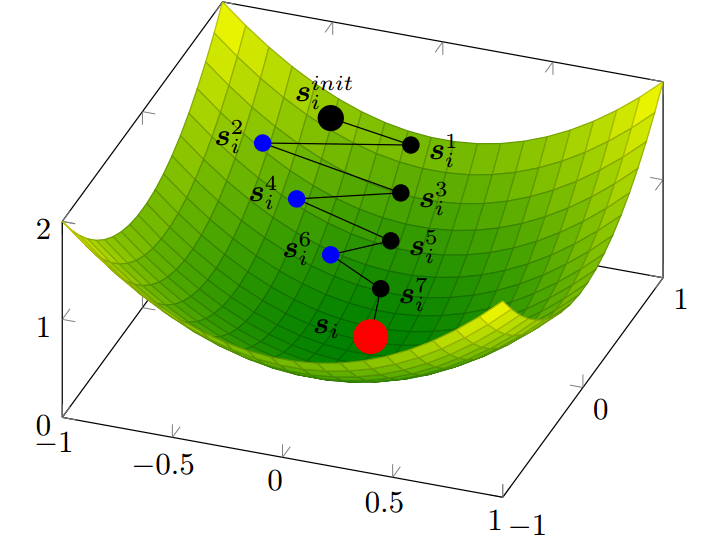}
    \caption{Scheme of the snapshots selection for $\Delta = 2$: black dots are discarded intermediate solutions, blue dots are saved intermediate solutions while the red dot represents the final solution.}
    \label{fig:snapSelection}
\end{figure}

Let  $\mathbb{P} = \{ \pi_1, \ldots, \pi_{N_{\pi}} \}$ be the training parameters set. For every parameter $\pi_i \in \mathbb{P}$, the full-order problem can be solved to obtain the corresponding solution $\bm{s}_i$. All these offline solutions are then stored in the snapshots matrix:
\begin{equation*}
    \bm{S} = \begin{bmatrix}
    s_{1_1} & s_{2_1} & \dotsc & s_{{N_{\pi}}_1}\\
    \vdots & \vdots & \vdots & \vdots\\
    s_{1_{N_h}} & s_{2_{N_h}} & \dotsc & s_{{N_{\pi}}_{N_h}}\\
    \end{bmatrix}.
\end{equation*}
In our case we want to construct an online solver able to mimic the offline convergence dynamics. For this reason the use of a monolithic (\textit{non-segregated}) approach for the reduced problem is not a good choice as the \textit{offline} solutions are obtained relying on a segregated solver; also at the \textit{online} level a segregated strategy has to be applied to obtain solutions which are as consistent as possible. For a discussion on a similar consistent approach in the context of explicit time integration schemes the reader is referred to \cite{StarSanderseStabileRozzaDegroote2020}.
To obtain an algorithm able to properly follow the behavior of the high fidelity algorithm, 
the set of snapshots is enriched by adding a certain amount of intermediate solutions $\bm{s}_i^j$ obtained during the offline iterations. The distance between exported intermediate solutions is set to $\Delta$ as shown in \autoref{fig:snapSelection}. Since the solution fields
during these iterations vary a lot, from the first attempt for the variables to last resolution, the
information contained into the converged snapshots is not sufficient to ensure the correct reduced
reconstruction of the path to the global minimum for \autoref{eq:COMP}.
By adding some non-physical solutions to the snapshots matrix, which is what is happening by inserting non-converged fields, we are somehow polluting the physical content but the convergence properties of the algorithm are \textit{quite acceptable} in any case. To reach a balance between convergence and reliability, $\Delta$ can be varied and the total amount $N_{int}$ of selected intermediate solutions can be modified. The new snapshots matrix then reads:
\begin{equation*}
    \bm{S} = \left[ \bm{s}_1^1, \bm{s}_1^2, \ldots, \bm{s}_1^{N_{int}}, \bm{s}_1, \dots, \bm{s}_{N_{\pi}}^1, \bm{s}_{N_{\pi}}^2, \cdots, \bm{s}_{N_{\pi}}^{N_{int}}, \bm{s}_{N_{\pi}}\right],
\end{equation*}
where $\bm{s}_i^j$ is the solution obtained at the $(j \, \Delta)$-th iteration for the $i$-th offline parameter.
In a POD-Galerkin approach, the reduced order solution $\bm{s}^r$ is obtained as a linear combination of some pre-calculated basis functions $\bm{\xi}$:
\begin{equation}
    \bm{s}^r(\bm{x}, \pi) = \sum_{i=1}^{N_r} \beta_i(\pi) \bm{\xi}_i(\bm{x}),
\end{equation}
where $N_r < N_{\pi}$ is the number of basis functions to be used for the reconstruction and the $\beta_i$ are the coefficients depending only on the parameter representing the reduced solution.

Once provided a certain amount $N_t$ of high fidelity solutions, with $N_t > N_{\pi}$ because of the intermediate snapshots, the best reduced order model we can get is the one able to fully reproduce the training offline solutions with no error with respect to it. Of course this is not achievable but we would like the $L^2$ norm of the error $E_{ROM}$ between all the offline solutions and the respective online ones to be as low as possible:
\begin{equation*}
    E_{ROM} = \sum_{i=1}^{N_t} \lvert \lvert \bm{s}^{ROM}_i - \bm{s}_i \rvert \rvert_{L^2} = \sum_{i=1}^{N_t} \Bigg|\Bigg| \sum_{j=1}^{N_r} \beta_j(\pi) \bm{\xi}_j(\bm{x}) - \bm{s}_i \Bigg| \Bigg|_{L^2}.
\end{equation*}
It is well known (see, e.g., \cite{kunisch2002galerkin}) that the basis functions best performing in this sense are the ones obtained through a Proper Orthogonal Decomposition (POD) applied to the snapshots matrix $\bm{S}$. The eigen problem 
\begin{equation*}
    \bm{C} \bm{V} = \bm{V} \bm{\lambda},
\end{equation*}
has to be resolved, where $\bm{C} \in \mathbb{R}^{N_t \times N_t}$ is the correlation matrix containing all the inner products in the form $(\bm{s}_i, \bm{s}_j)_{L^2(\Omega)}$.  $\bm{V} \in \mathbb{R}^{N_t \times N_t}$ is the matrix containing its eigenvectors while $\bm{\lambda}$ is the diagonal matrix containing the eigenvalues.

The basis functions are then constructed as just a linear combination of the snapshots contained in $\bm{S}$:
\begin{equation*}
    \bm{\xi}_i(\bm{x}) = \frac{1}{{N_t\sqrt{\lambda_i}}} \sum_{j=1}^{N_t} \bm{V}_{ji} \bm{s}_j (\bm{x}).
\end{equation*}
The basis functions matrix is then defined as:
\begin{equation*}
    \bm{\Xi} = \left[ \bm{\xi}_1, \dotsb, \bm{\xi}_{N_r} \right] \in \mathcal{R}^{N_h \times N_r} .
\end{equation*}

The interested reader may refer to \cite{quarteroni2015reduced,benner2017model,hesthaven2016certified} for a detailed explanation regarding POD approaches.

%% file: sections/compRedSimple.tex
\subsection{A segregated reduced approach for compressible flows}\label{sec:redComp}

\autoref{sec:pod} introduces the intention of this work: the construction of a new segregated approach for compressible flows. By the procedure explained in the previous section, it is possible to construct three different snapshots matrices for pressure, velocity and energy respectively: $\bm{S}_{p} \in \mathbb{R}^{N_h \times N_t}$, $\bm{S}_{u} \in \mathbb{R}^{d \, N_h \times N_t}$ and $\bm{S}_{e} \in \mathbb{R}^{N_h \times N_t}$ being $d$ the spacial dimension of the problem. They are employed to extract three basis function sets so that all the different variables of \autoref{eq:FANS} can be written into their reduced form:
\begin{eqnarray*}
    \overline{p}^r = \sum_{i=1}^{N_p} a_i (\pi) \varphi_i (\bm{x}) = \bm{\Phi} \bm{a} \in \mathbb{Q}_r, \\
     \tilde{\bm{u}}^r = \sum_{i=1}^{N_u} b_i (\pi) \bm{\xi}_i (\bm{x}) = \bm{\Xi} \bm{b} \in \mathbb{V}_r,\\
     \tilde{e}^r = \sum_{i=1}^{N_e} c_i (\pi) \theta_i (\bm{x}) = \bm{\Theta} \bm{c} \in \mathbb{E}_r,
\end{eqnarray*}
where $\mathbb{Q}_r = \text{span} \{\varphi_i\}_{i=1}^{N_p} \subset \mathbb{Q}_h$, $\mathbb{V}_r = \text{span} \{\psi_i\}_{i=1}^{N_u} \subset \mathbb{V}_h$ and $\mathbb{E}_r = \text{span} \{\theta_i\}_{i=1}^{N_e}  \subset \mathbb{V}_h$ are the reduced spaces, $\bm{a} \in \mathbb{R}^{N_p}$, $\bm{b} \in \mathbb{R}^{N_u}$ and $\bm{c} \in \mathbb{R}^{N_e}$ are the vectors containing the coefficients $a_i$, $b_i$ and $c_i$ depending only on the parameter value related to pressure, velocity and energy respectively, $\bm{\Phi} \in \mathbb{R}^{N_h \times N_p}$, $\bm{\Psi} \in \mathbb{R}^{d \, N_h \times N_u}$ and $\bm{\Theta} \in \mathbb{R}^{N_h \times N_e}$ are the matrices containing the modal basis functions $\varphi_i$, $\bm{\psi}_i$ and $\theta_i$ related to pressure, velocity and energy respectively while $N_p < N_t$, $N_u < N_t$ and $N_e < N_t$ are the numbers of modal basis functions selected for pressure, velocity and energy to reconstruct their reduced solutions. This means that the procedure explained in \autoref{sec:pod} has to be applied three times to the three different solutions sets $\bm{S}_{p}$, $\bm{S}_{u}$ and $\bm{S}_{e}$.
\begin{algorithm*}[ht]
\caption{The reduced-order SIMPLE algorithm}
\label{alg:romSimple}
\hspace*{\algorithmicindent} \textbf{Input:} first attempt reduced pressure, velocity and energy coefficients $\bm{a}^\star$, $\bm{b}^\star$ and $\bm{c}^\star$; modal basis functions matrices for pressure, velocity and energy $\Phi$, $\Xi$ and $\Theta$\\
\hspace*{\algorithmicindent} \textbf{Output:} reduced pressure, velocity and energy fields $\overline{p}_r$, $\tilde{\bm{u}}_r$ and $\tilde{e}_r$
\begin{algorithmic}[1]
  \State From $\bm{a}^\star$, $\bm{b}^\star$ and $\bm{c}^\star$, reconstruct reduced fields $\overline{p}^\star$, $\tilde{\bm{u}}^\star$ and $\tilde{e}^\star$:~~ $\overline{p}^\star = \Phi \bm{a}^\star {,} \hspace{0.5cm} \tilde{\bm{u}}^\star = \Xi \bm{b}^\star {,} \hspace{0.5cm} \tilde{e}^\star = \Theta \bm{c}^\star$ \;
  \State Evaluate the eddy viscosity field $\mu_t$ with the \textbf{neural network};
  \State Momentum predictor step : assemble \autoref{eq:mMomEq}, \textbf{relax} it employing prescribed under-relaxation factor $\alpha_u$, project it over the velocity basis functions $\psi_i$ and solve it to obtain new reduced velocity coefficients vector $\bm{b}^{\star\star}$;
  \State Reconstruct the new reduced velocity $\tilde{\bm{u}}^{\star\star}$ and calculate the off-diagonal component $\bm{H}(\tilde{\bm{u}}^{\star\star})$;
  \State Energy equation step : assemble \autoref{eq:mEnEq}, relax it employing prescribed under-relaxation factor $\alpha_e$, project it over the energy basis functions $\theta_i$ and solve it to obtain new reduced energy coefficients vector $\bm{c}^{\star\star}$;
  \State Reconstruct the new reduced energy $\tilde{e}^{\star\star}$;
  \State Calculate both density $\overline{\rho}^{\star\star}$ and temperature $\tilde{T}^{\star\star}$ fields starting from $\overline{p}^\star$, $\tilde{\bm{u}}^{\star\star}$ and $\tilde{e}^{\star\star}$ by the use of the state equation;
  \State Pressure correction step: assemble \autoref{eq:mPresEq}, project it over the pressure basis functions $\varphi_i$ to get new reduced pressure coefficients $\bm{a}^{\star\star}$; then correct the velocity explicitly after having reconstructed the new pressure $\overline{p}^{\star\star}$;
  \State Relax the pressure field with the prescribed under-relaxation factor $\alpha_p$. The under-relaxed field is called $\overline{p}^{ur}$;
  \If {convergence}
      \State $\overline{p}_r = \overline{p}^{ur}$, $\tilde{\bm{u}}_r = \tilde{\bm{u}}^{\star\star}$ and $\tilde{e}_r = \tilde{e}^{\star\star}$ 
  \Else 
      \State Assemble the conservative face fluxes $F_{f}$:
    	$F_{f} = \tilde{\bm{u}}_{f} \cdot \bm S_{f}$ \;
      \State set $\overline{p}^\star = \overline{p}^{ur}$, $\tilde{\bm{u}}^{\star} = \tilde{\bm{u}}^{\star\star}$ and $\tilde{e}^{\star} = \tilde{e}^{\star\star}$;
      \State iterate from step 1.;
  \EndIf
\end{algorithmic}
\end{algorithm*}
Here, relax is given by:
\begin{equation}
    Q^n = Q^{n-1} + \alpha(Q^{n*} - Q^{n-1}).
\end{equation}
Where $\alpha$ is the factor that defines the relaxation such that:
\begin{itemize}
    \item $\alpha < 1$ means under-relaxation. This will slow down the convergence rate but increase the stability.
    \item $\alpha = 0$  means no relaxation at all. The predicted value of $Q$ is simply used.
    \item  $\alpha > 1$ means over-relaxation. It can sometimes be used to accelerate the convergence rate but will decrease stability.
\end{itemize}
$n$ refers to the new, used value of $Q$, $n-1$  refers to the previous value of $Q$, and $n*$  refers to the new predicted value of $Q$.

%% file: sections/turbTreatment.tex
\subsection{Turbulence treatment}\label{sec:turbTr}
\begin{figure}[htbp]
    \centering
    \tikzset{%
      every neuron input/.style={
        circle,
        draw,
        minimum size=0.5cm,
        fill=green!50
      },
      every neuron input2/.style={
        circle,
        draw,
        minimum size=0.5cm,
        fill=yellow!50
      },
       every neuron hidden/.style={
        circle,
        draw,
        minimum size=0.5cm,
        fill=blue!50
      },
       every neuron output/.style={
        circle,
        draw,
        minimum size=0.5cm,
        fill=red!50
      },
      neuron missing/.style={
        draw=none, 
        fill=none,
        scale=4,
        text height=0.333cm,
        execute at begin node=\color{black}$\vdots$
      },
       layer missing/.style={
        draw=none, 
        scale=4,
        text height=0.333cm,
        execute at begin node=\color{black}$\dots$
      },
    }

    \begin{tikzpicture}[x=1.5cm, y=1.5cm, >=stealth, font=\small, scale=0.6]
    
    \foreach \i [count=\y] in {1, missing, 2, 3, missing, 4}
     \ifthenelse {\y=4 \OR \y=6}{\node [every neuron input2/.try, neuron \i/.try] (input-\i) at (0,2.5-\y) {}}{\node [every neuron input/.try, neuron \i/.try] (input-\i) at (0,2.5-\y) {}};
    
    \foreach \m [count=\y] in {1, 2, missing,3}
      \node [every neuron hidden/.try, neuron \m/.try ] (hidden1-\m) at (2,2-\y) {};
     
    \foreach \m [count=\y] in {1, missing, 2, 3}
      \node [every neuron hidden/.try, neuron \m/.try ] (hidden2-\m) at (4,2-\y) {};
    
    \foreach \m [count=\y] in {1,missing,2}
      \node [every neuron output/.try, neuron \m/.try ] (output-\m) at (6,1.5-\y) {};
    

    \draw [<-] (input-3) -- ++(-1,0)
        node [above, midway] {$b_1$};
    \draw [<-] (input-4) -- ++(-1,0)
        node [above, midway] {$b_{N_{u}}$};
    \draw [<-] (input-1) -- ++(-1,0)
        node [above, midway] {$\pi_1$};
    \draw [<-] (input-2) -- ++(-1,0)
        node [above, midway] {$\pi_{N_{\pi}}$};
    
    \foreach \l [count=\i] in {1,N_{\mu_t}}
      \draw [->] (output-\i) -- ++(1,0)
        node [above, midway] {$m_{\l}$};
    
    \foreach \i in {1,...,4}
      \foreach \j in {1,...,3}
        \draw [->] (input-\i) -- (hidden1-\j);
    
    \foreach \i in {1,...,3}
      \foreach \j in {1,...,3}
        \draw [->] (hidden1-\i) -- (hidden2-\j);
    
    \foreach \i in {1,...,3}
      \foreach \j in {1,...,2}
        \draw [->] (hidden2-\i) -- (output-\j);
    
    \foreach \l [count=\x from 0] in {Input \\ layer, Hidden \\ layer  $f_{1}$}
    \node [align=center, above] at (\x*2,2) {\l};
    \node [align=center, above] at (4,2) {Hidden \\ layer $f_2$};
    \node [align=center, above] at (6,2) {Output \\ layer $f_{out}$};
      
    \end{tikzpicture}
    \caption{Schematic perspective of a fully connected neural network composed by an input layer, two hidden layers and an output layer, linking parameters $\pi_i$ and reduced velocity coefficients $b_i$ to reduced eddy viscosity coefficients $m_i$, being $N_{\pi}$ the number of parameters possibly existing in the problem.}
    \label{fig:nn_nut}
\end{figure}
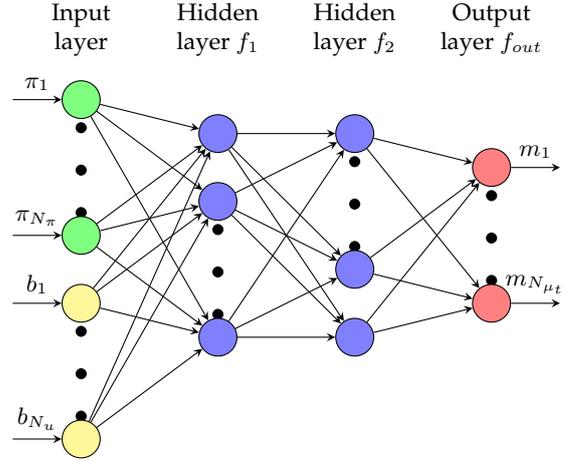
In this work some assumptions were taken in \autoref{sec:compNS} and \autoref{sec:fvDisc} leading to a simplified FANS system, \autoref{eq:rhosimple}. Turbulence effects in \autoref{eq:rhosimple} are all due to the presence of the eddy viscosity field $\mu_t$. A technique has to be selected to model the eddy viscosity. Within this scope, many different approaches are possible \cite{siddiqui2020numerical,wilcox1976complete,alfonsi2009reynolds,sirovich1987turbulence}.

To make our architecture as independent as possible on the turbulence model used during the offline phase to evaluate the $\mu_t$ field, this study combines a classical POD-Galerkin approach for what concerns the physical variables $\overline{p}, \tilde{\bm{u}}$ and $\tilde{e}$ together with a data driven scheme for what concerns the eddy viscosity evaluation in the Boussinesq hypothesis \cite{schmitt2007boussinesq}.

Let us imagine to approximate the eddy viscosity field similarly to what has been done for all the other variables:
\begin{equation*}
    \mu_{t_r} = \sum_{i=1}^{N_{\mu_t}} m_i (\pi) \eta_i (\bm{x}),
\end{equation*}
where $N_{\mu_t}$ is the number of basis functions selected to reconstruct the eddy viscosity field, $m_i$ are the coefficients depending only on the position $\bm{x}$ while $\eta_i$ are the $\mu_t$ basis functions depending only on the parameter. During the offline phase, together with all the other saved solutions, also the eddy viscosity fields are exported and stored. Those snapshots are then collected into the $\bm{S}_{\mu_t}$ matrix and used, as explained in \ref{sec:pod}, to obtain the requested basis functions.
For what concerns the spacial coefficients, they are evaluated through a Neural Network (NN) scheme linking the parameters of the problem $\pi_i$ and the reduced velocity coefficients $b_i$ to the $m_i$. In fact it is well known that, no matter what turbulence model is employed, the eddy viscosity $\mu_t$ depends on the velocity field but, especially for geometrically parametrized problems, it also depends on the parameter itself. The reduced problem is thus completely independent on the choice of the turbulence model and \textit{step 2} into \autoref{alg:romSimple} can be performed in an efficient way. This would not have been the case if turbulence equations were projected: in case there was the necessity of changing the adopted turbulence model, all the architecture had to be modified.

In this work,  we selected a fully connected Neural Network composed by an input layer, two hidden layers and an output layer. The input vector $\bm{z}$ and output vector $\bm{m}$ are defined as mentioned before:
\begin{equation*}
    \bm{z}^T=
    \begin{bmatrix}
    \pi_1,
    \cdots,
    \pi_{N_{\pi}}, 
    b_1,
    \cdots, 
    b_{N_u}
    \end{bmatrix}, \hspace{0.05cm}
    \bm{m}^T=
    \begin{bmatrix}
    m_1,
    \cdots,
    m_{N_{\mu_t}},
    \end{bmatrix}.
\end{equation*}
It is clear that the Neural Network has to be trained in some way. To this scope the snapshots contained into $\bm{S}_{\mu_t}$ are projected over their own basis functions $\eta_i$ to obtain the set of real coefficients $\{\bm{m}_i\}_{i=1}^{N_t}$. They can be compared with the NN estimated coefficients $\{\tilde{\bm{m}}_i\}_{i=1}^{N_t}$ into a loss function to target the training procedure. The loss function $\ell$ we adopted is a widely used quadratic one:
\begin{equation*}
    \ell = ||\bm{m} - \tilde{\bm{m}}||_{L^2}.
\end{equation*}
The quantity $\mathcal{L}$ to be minimized during the training of the network is the sum of the loss function evaluated for all the different snapshots:
\begin{equation*}
    \mathcal{L} = \sum_{i=1}^{N_t} ||\bm{m}_i - \tilde{\bm{m}}_i||_{L^2}.
\end{equation*}
The coefficients estimated by the network can be written as:
\begin{equation*}
    \tilde{\bm{m}} = \bm{\textit{f}}_{out} \left( \bm{W}_{out} \, \bm{\textit{f}}_2\left( \bm{W_2} \bm{\textit{f}}_1 \left( \bm{W_1} \bm{x} + \bm{b}_1\right) + \bm{b}_2\right) + \bm{b}_{out} \right),
\end{equation*}
where $\bm{\textit{f}}_1$, $\bm{\textit{f}}_2$ and $\bm{\textit{f}}_{out}$ are the activation functions, $\bm{W_1}$, $\bm{W_2}$ and $\bm{W_{out}}$ are the weights while $\bm{b}_1$, $\bm{b}_2$ and $\bm{b}_{out}$ are the biases, related to the first and the second hidden layers and to the output layer respectively. For the hidden layers the best performing activation function appears to be the hyperbolic tangent while the output layer has been simply implemented as a linear combination of the received data. The previous formula can then be simplified as follows:
\begin{equation*}
    \tilde{\bm{m}} = \bm{W}_{out} \, \tanh\left( \bm{W_2} \, \tanh \left( \bm{W_1} \bm{x} + \bm{b}_1\right) + \bm{b}_2\right) + \bm{b}_{out},
\end{equation*}
where $\tanh(\bm{y})^T = \begin{bmatrix} \tanh(y_1),  \cdots,  \tanh(y_{dim_y}) \end{bmatrix}$,  being $\bm{y} = [y_1, \hdots, y_{dim_y}]$ a generic vector quantity.

%% file: sections/physical.tex
\subsection{Physical parametrization test case}\label{sec:phy}

The first test case we present in this work is a physically parameterized external flow: a NACA0012 airfoil is immersed into a fluid with variable viscosity $\mu$. The unperturbed velocity is fixed and is equal to $\tilde{\boldsymbol{u}}_{inlet}=[ 250, 0, 0]^T$\si{m\per\second} while the chord of the airfoil is equal to one. As already said, the viscosity can vary so that $\mu \in [10^{-5}, 10^{-2}]$. The speed of sound at the inlet can easily be evaluated by taking into consideration the thermophysical properties of the gas we are working with. We consider perfect gasses. Thus the specific heat transfer at constant pressure is sufficient to evaluate $\gamma = \frac{C_p}{C_v} = \frac{C_p}{C_p-R}$ where $C_p=1005\si{\joule\kilogram\per\kelvin}$ while $R=8,314 \si{\joule\mol\per\kelvin}$ is the constant for perfect gasses. We suppose our airfoil to move into air so that $M=28,9$ \si{\g\per\mol} where $M$ stands for the molar weight. Temperature is fixed at $T=298\si{\K}$. Collecting all these data together, we end up with
$C = \sqrt{\frac{\gamma R T}{M}} = 341.17 \si{\m\per\s}.$
This means that at the inlet the Mach number can be calculated as 
$$\text{Mach}=\frac{\tilde{\bm{u}}_{inlet}}{C} \simeq 0.73.$$ 
For this test case, consequently, a compressible treatment for the flow is needed since we are approaching the \textit{Transonic regime} and compressible effects are pretty significant.
At the inlet, pressure is fixed to $10^5$ Pa. Then the Reynolds number can be evaluated as 
$$
\text{Re}=\frac{\rho L \tilde{\bm{u}}_{inlet}}{\mu}=\frac{p L \tilde{\bm{u}}_{inlet} M}{\mu R T}.
$$
The resulting Reynolds number is then $\text{Re}\in 2.92 \times \left[10^4, 10^7 \right]$, which clearly requires treatment for turbulence since the system is operating in a fully turbulent regime.

For the offline phase, $50$ random values have been selected: $\pi_i \in [10^{-5}, 10^{-2}] \ \text{for} \ i=1,\ldots,50$ where $[\pi_1, \ldots, \pi_{50}]=\mathbb{P}$. Full-order eddy viscosity is calculated by the resolution of a $k-\omega$ turbulence model \cite{wilcox1998turbulence}.
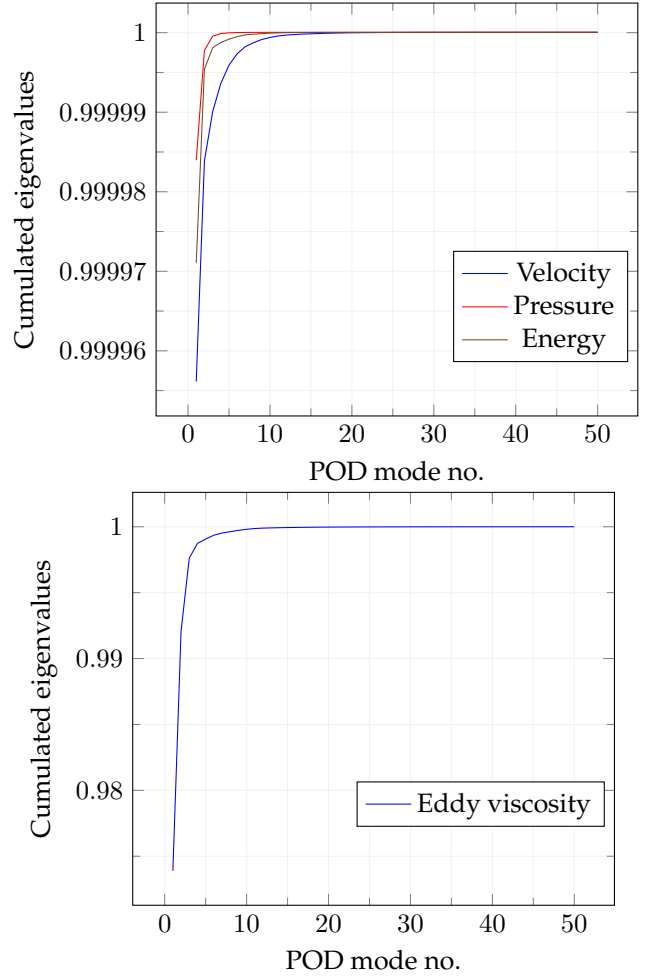
\begin{figure}[h]
    \pgfplotsset{scaled y ticks=false}
    \centering
    \begin{tikzpicture}
        \begin{axis}[
            name=plot1,
            xlabel={POD mode no.},
            ylabel={Cumulative eigenvalues},
            grid=both,
            grid style={line width=.1pt, draw=gray!10},
            minor tick num=1,
            y tick label style={
            /pgf/number format/.cd,
            fixed,
            precision=5,
            /tikz/.cd
            },
            legend style={at={(0.98,0.4)},anchor=north east},
            height=0.8*0.49\textwidth,
            width=0.44*\textwidth, 
        ]
        \addplot+[no marks]
            table[x={xAx}, y={eigsU}] {physicalData/eigs.dat};
            
        \addplot+[no marks]
            table[x={xAx}, y={eigsP}] {physicalData/eigs.dat};
            
        \addplot+[no marks]
            table[x={xAx}, y={eigsE}] {physicalData/eigs.dat};
        
        \legend{Velocity, Pressure, Energy};
        \end{axis}
        \end{tikzpicture}
        \begin{tikzpicture}
        \begin{axis}[
            name=plot2,
            at=(plot1.right of south east),
            anchor=left of south west,
            xlabel={POD mode no.},
            ylabel={Cumulative eigenvalues},
            legend style={at={(0.98,0.3)},anchor=north east},
            height=0.8*0.49\textwidth,
            width=0.44*\textwidth, 
            grid=both,
            grid style={line width=.1pt, draw=gray!10},
            minor tick num=1,
            y tick label style={
            /pgf/number format/.cd,
            fixed,
            precision=5,
            /tikz/.cd
            },
        ]

        \addplot+[no marks]
            table[x={xAx}, y={eigsNut}] {physicalData/eigs.dat};
        
        \legend{Eddy viscosity};
        \end{axis}
    \end{tikzpicture}
    \caption{Cumulative eigenvalues trends.}
    \label{fig:eigsPh}
\end{figure}
\autoref{fig:eigsPh} shows the trends of the \textit{cumulative} eigenvalues for velocity, pressure, energy, and eddy viscosity. As we may notice, by just considering a few modes for every variable, the amount of discarded information is pretty low. For this reason, just the first $20$ modal basis functions have been selected for velocity, pressure, and energy while $30$ modal basis functions are used to reconstruct the eddy viscosity field. This is due to the fact that analyzing \autoref{fig:eigsPh}, it is clear that a higher number of basis functions are needed in order to approach the unity in the cumulative eigenvalues plot.

For what concerns the neural network for the eddy viscosity coefficients, as explained in \autoref{sec:turbTr}, two hidden layers are present, the first one composed of $256$ neurons and the second one composed of $64$ neurons, resulting in a fully connected network where only $\tanh$ activation functions are used. Offline solutions, including the intermediate steps, are retained to train the network. 
\begin{figure}[h]
    \centering
    \begin{tikzpicture}
        \begin{semilogyaxis}[
            xlabel={Epochs},
            ylabel={Loss},
            legend style={at={(0.98,0.98)},anchor=north east},
            height=0.8*0.49\textwidth,
            width=0.49*\textwidth, 
            grid=both,
            grid style={line width=.1pt, draw=gray!10},
            minor tick num=1,
            x tick label style={
                               /pgf/number format/.cd,
                                    fixed,
                               /tikz/.cd
            },
            scaled x ticks=false
        ]
        \addplot+[no marks]
            table[x={xAxL}, y={trainLoss}] {physicalData/loss2030.dat};
            
        \addplot+[no marks]
            table[x={xAxL}, y={testLoss}] {physicalData/loss2030.dat};
        
        \legend{Train, Test};
        
        \end{semilogyaxis}
    \end{tikzpicture}
    \caption{Loss function decay for both train and test sets.}
    \label{fig:lossesPh}
\end{figure}
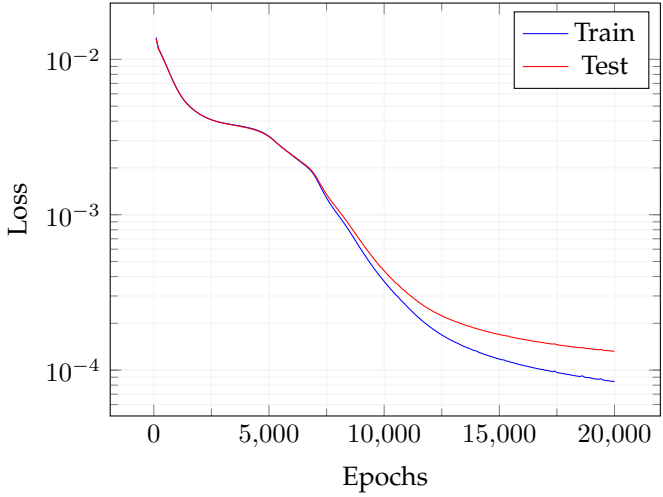
The training procedure is carried out in $2000$ epochs. $20$ new random offline solutions have been performed to obtain a testing set that was not correlated to the solutions used for the training stage. A mean squared error loss function is used to evaluate the reconstruction capability of the network for both training and testing sets. The decay behavior of both losses is depicted in \autoref{fig:lossesPh}. The training stage was stopped after $2 \times 10^3$ epochs to avoid over-fitting and the distance between test and train losses was starting to increase significantly. 
\begin{figure}[h]
    \centering
    \begin{tikzpicture}
        \begin{axis}[
            name=plot1,
            xlabel={Parameter no.},
            ylabel={$L^2$ relative error},
            legend style={at={(0.98,0.7)},anchor=north east},
                        grid=both,
            grid style={line width=.1pt, draw=gray!10},
            minor tick num=1,
            height=0.8*0.49\textwidth,
            width=0.49*\textwidth, 
        ]

        \addplot+[only marks]
            table[x={parameter}, y={errorU}] {physicalData/errors.dat};
        \addplot+[only marks]
            table[x={parameter}, y={errorP}] {physicalData/errors.dat};
        \addplot+[only marks]
            table[x={parameter}, y={errorE}] {physicalData/errors.dat};
        
        \legend{Velocity, Pressure, Energy};
        \end{axis}
            \end{tikzpicture}
            \begin{tikzpicture}

        \begin{axis}[
            name=plot2,
            at=(plot1.right of south east),
            anchor=left of south west,
            xlabel={Parameter no.},
            ylabel={$L^2$ relative error},
            legend style={at={(0.98,0.3)},anchor=north east},
                        grid=both,
            grid style={line width=.1pt, draw=gray!10},
            minor tick num=1,
            height=0.8*0.49\textwidth,
            width=0.49*\textwidth, 
        ]

        \addplot+[only marks]
            table[x={parameter}, y={errorNut}] {physicalData/errors.dat};
        
        \legend{Eddy viscosity};
        \end{axis}
    \end{tikzpicture}
    \caption{$L^2$ norm relative errors.}
    \label{fig:errorsPh}
\end{figure}
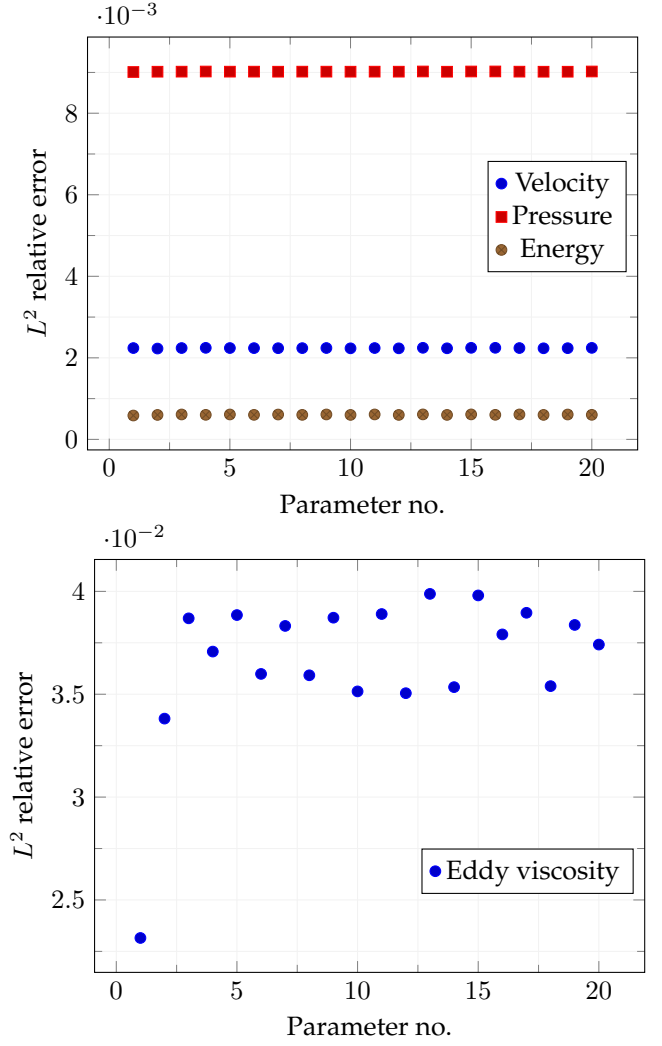
\autoref{fig:errorsPh}, left, shows the $L^2$ norm relative errors for all the different parameters in the online set concerning velocity, pressure, and internal energy. \autoref{fig:errorsPh}, right, shows the $L^2$ norm relative error for the eddy viscosity between full order and reduced order for the whole online parameter set. As we may notice, even if the order of magnitude of the $\nu_t$ error is equal to $10^{-2}$, it is sufficient to ensure a lower error for the quantities of interest, i.e. velocity, pressure, and energy. By this observation we are allowed to employ such a small neural network which is not compromising the computational cost, still ensuring good performances.  
\begin{figure}
    \centering
    \begin{tabular}{cc}
    \includegraphics[width=0.32 \textwidth]{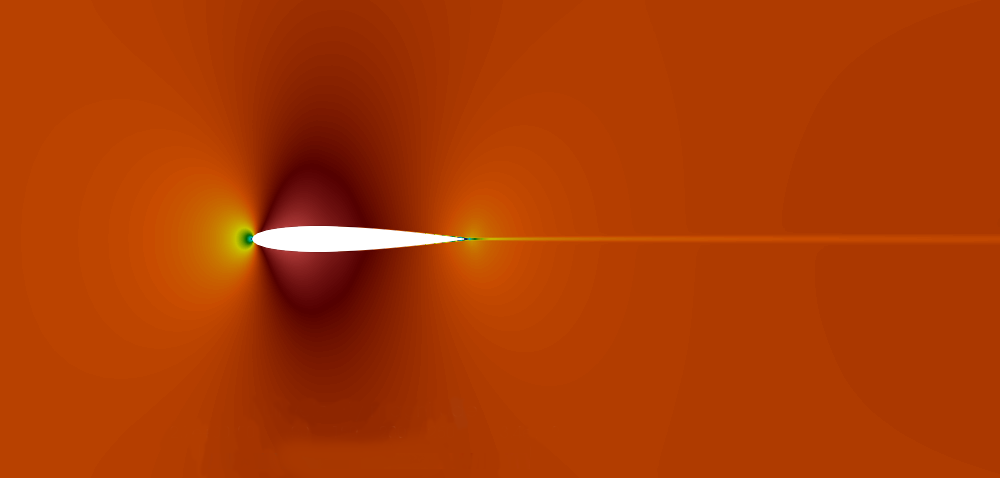}
    \includegraphics[width=0.32 \textwidth]{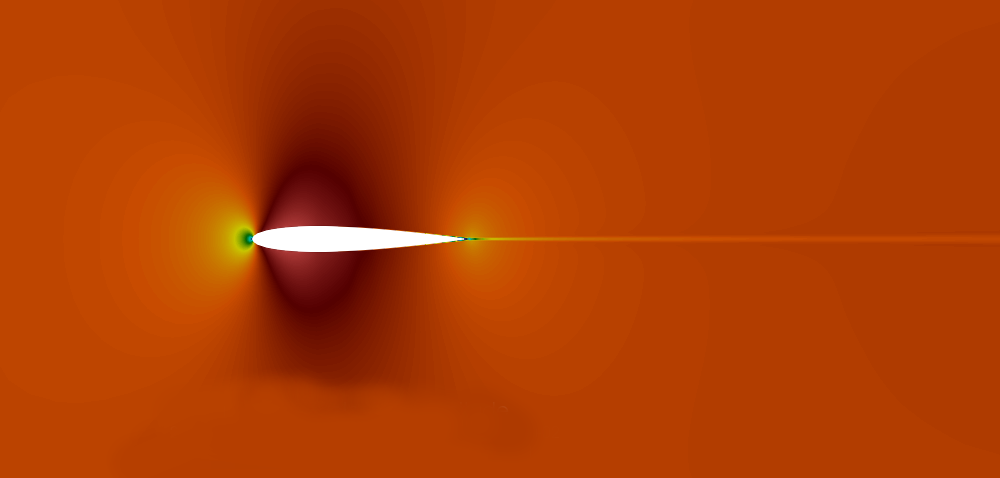}\\
    \includegraphics[width=0.32 \textwidth]{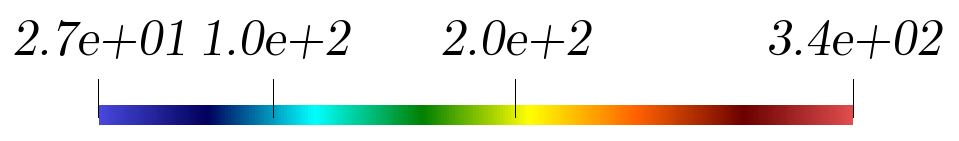}
    \end{tabular}
    \begin{tabular}{cc}
    \includegraphics[width=0.32 \textwidth]{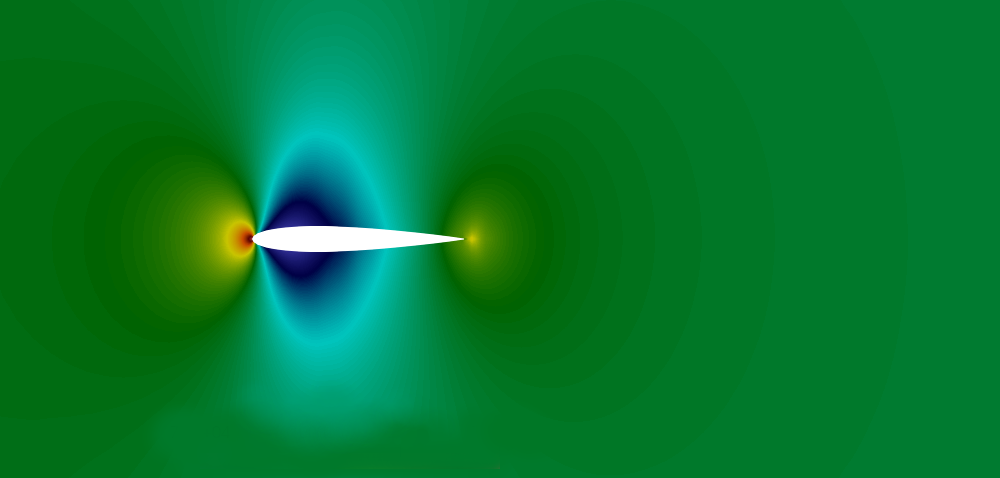}
    \includegraphics[width=0.32 \textwidth]{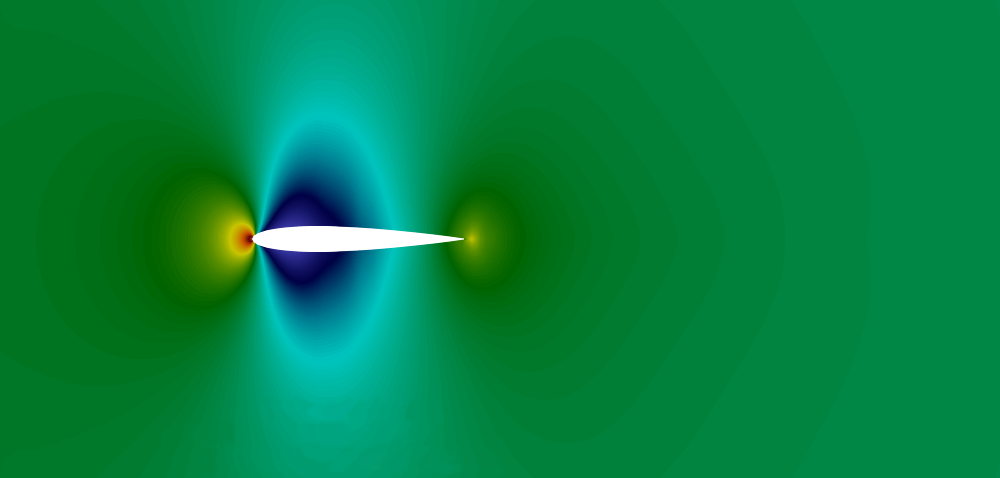}\\ 
    \includegraphics[width=0.32 \textwidth]{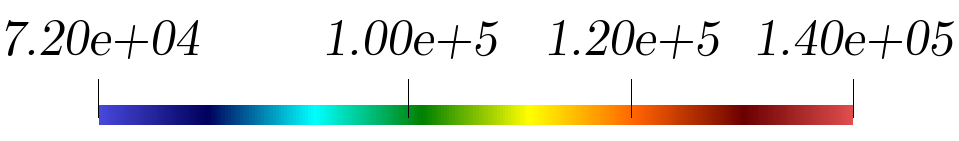}
    \end{tabular}
    \begin{tabular}{cc}
    \includegraphics[width=0.32 \textwidth]{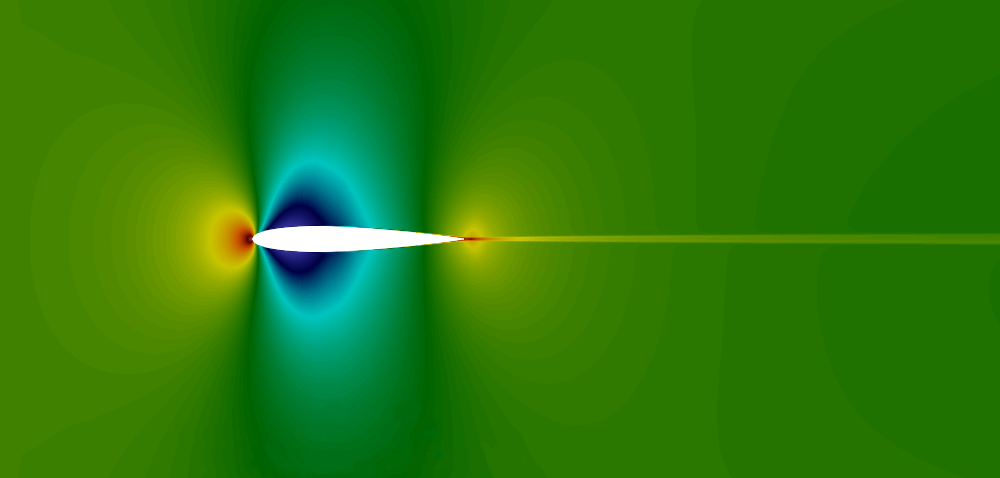}       
    \includegraphics[width=0.32 \textwidth]{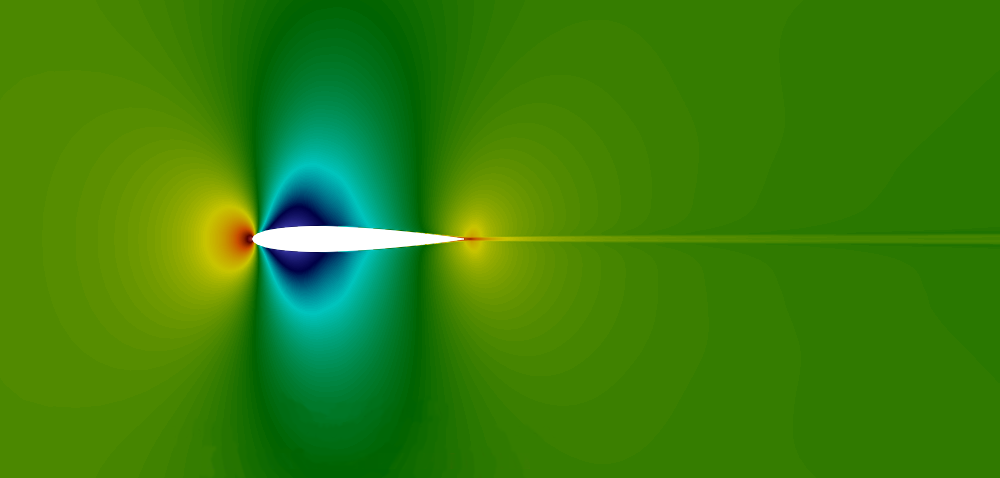} \\   
    \includegraphics[width=0.32 \textwidth]{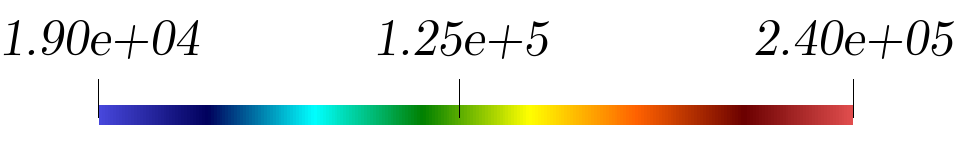}
    \end{tabular}
    \caption{Comparison between full-order (first column) and reduced-order (second column) solutions: \textit{velocity} magnitude (first row), \textit{pressure} (second row) and \textit{energy} (third row). These fields refer to the resolution of the problem for $\pi = \mu = 0.21 \times 10^{-3}$ which has been selected as a random value in the online parameter set.}
    \label{fig:physicalFields}
\end{figure}
\begin{figure}
    \centering
    \includegraphics[width=0.32 \textwidth]{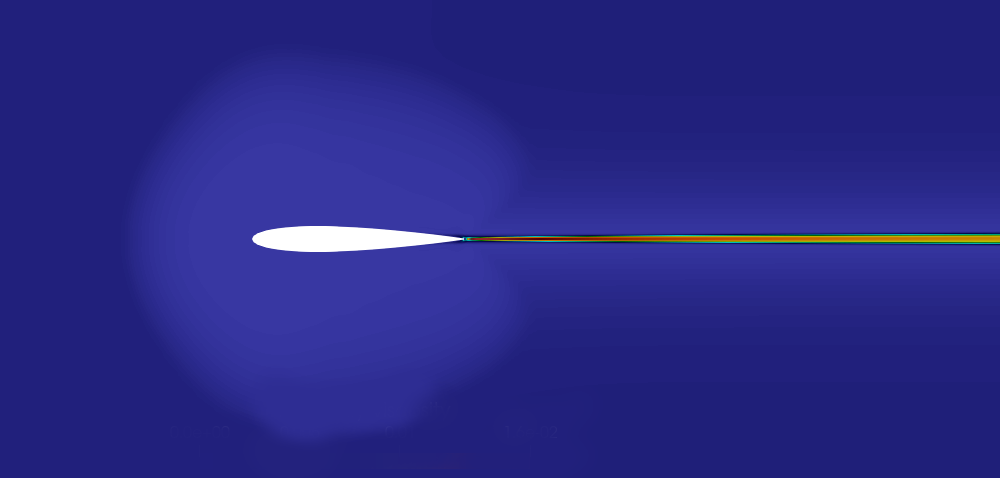}
    \includegraphics[width=0.32 \textwidth]{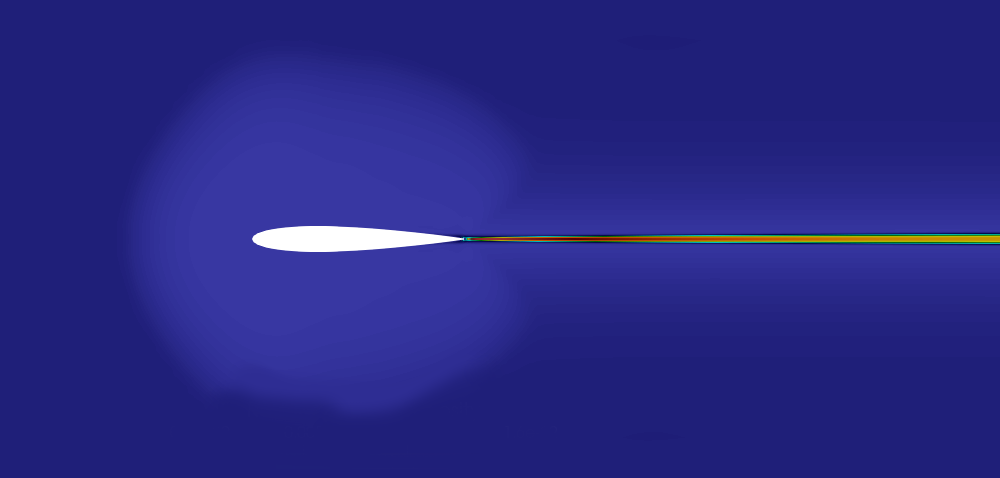}\\
    \includegraphics[width=0.32 \textwidth]{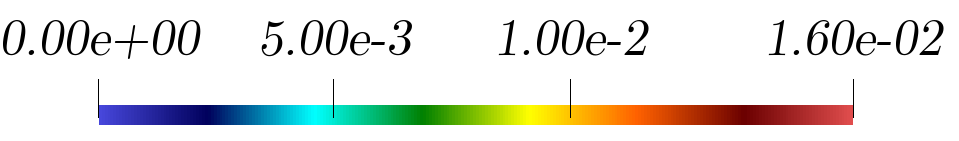}
    \caption{Comparison between full-order (left picture) and reduced-order (right picture)  for the eddy viscosity solutions. These fields refer to the resolution of the problem for $\pi = \mu = 0.21 \times 10^{-3}$ which has been selected as a random value in the online parameter set.}
    \label{fig:physicalNuts}
\end{figure}
In \autoref{fig:physicalFields} and \autoref{fig:physicalNuts} a comparison between full-order and reduced-order solutions is depicted, for a random value of the parameter, included in the online set. By analyzing the depicted fields, full-order and reduced-order solutions appear to be very similar, and the most important areas in the domain, i.e. the zone surrounding the airfoil together with the wake created by the body, are well reconstructed.

%% file: sections/geometrical.tex
\subsection{Geometrical parametrization test case}\label{sec:geom}
\begin{figure}[!htb]
    \centering
    \includegraphics[width=0.65\textwidth]{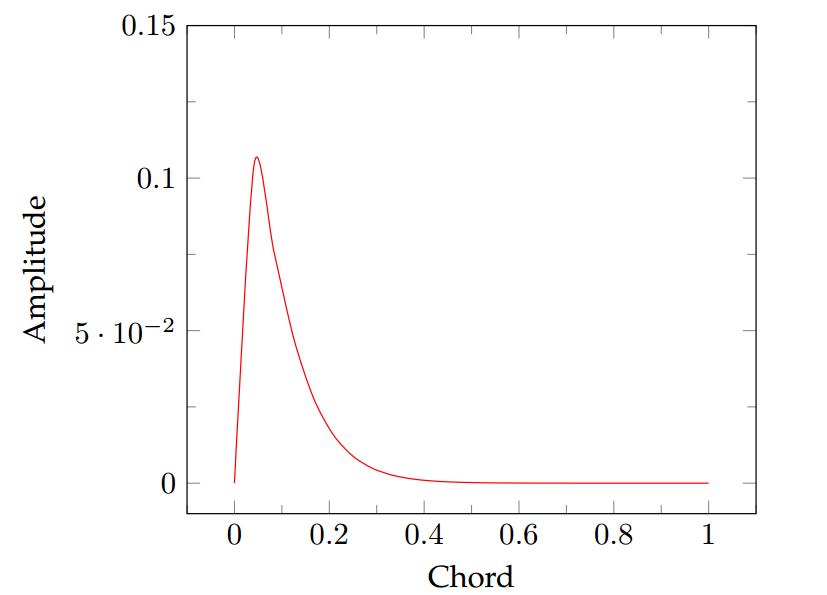}
    \caption{Shape of the employed bump function.}
    \label{fig:bump}
\end{figure}

This section presents the second test case, focused on a geometrically parameterized problem. The shape of the airfoil used into \autoref{sec:phy} is modified by the use of a bump function. In particular, the foil is divided in a top and a bottom part by the chord. The bump function depicted in \autoref{fig:bump} is added to the top and subtracted to the bottom surface, premultiplied by two different amplitude scalar factors: every solution is parameterised uniquely by two different scalar values. We used the same thermophysical properties used for \autoref{sec:phy} but the dynamic viscosity is fixed and equal to $1.74 \times 10^{-5} Pa \, s$. Moreover, the inlet velocity has been slightly decreased since the random modification of the geometry may lead to high curvature areas where the flow could eventually become supersonic: $\tilde{\bm{u}}_{inlet}=\begin{bmatrix} 170, 0, 0 \end{bmatrix}^T \si{\m\per\s}$. This means that the Mach number at the inlet is now around $0.5$.
For the offline phase, $50$ random values have been selected: $\pi_{top_i}, \pi_{bottom_i} \in [0,0.1]$ for $i= 1,...,50$ where
$\mathbb{P} = \{(\pi_{top_i}, \pi_{bottom_i})  \}_{i=1}^{50}$.
Full-order  eddy  viscosity  is  calculated  by  the  resolution  of  a $k-\omega$ turbulence model \cite{wilcox1998turbulence}.

The general POD approach described in \autoref{sec:pod} is not directly applicable to a geometrical parametrization problem since the $L^2$-norm used for the inner products is not well defined in case of multiple different domains. The mesh in our case is moved thanks to a Radial Basis Functions (RBF) algorithm where the points on the moving boundaries are displaced by the application of the desired law and their displacements are used as boundary conditions for an interpolation procedure, performed in order to move all the remaining points of the grid. The interested reader may find a deeper explanation of this technique in \cite{de2007mesh} or some applications in \cite{stabile2020efficient} and \cite{aria2021}. By exploiting the aforementioned method, the mesh is modified for each offline solution. To take into account the fact that all the snapshots are defined over a different mesh, the grid is taken back to its undeformed state before starting the POD procedure: the mass matrix we consider to evaluate the norms is then the reference unperturbed one. 
\begin{figure}[!htb]
    \pgfplotsset{scaled y ticks=false}
    \centering
    \begin{tikzpicture}
        \begin{axis}[
            name=plot1,
            xlabel={POD mode no.},
            ylabel={Cumulative eigenvalues},
            grid=both,
            grid style={line width=.1pt, draw=gray!10},
            minor tick num=1,
            height=0.49*0.8*\textwidth,
            width=0.44*\textwidth,
            y tick label style={
            /pgf/number format/.cd,
            fixed,
            precision=5,
            /tikz/.cd
            },
            legend style={at={(0.98,0.4)},anchor=north east},
        ]
        \addplot+[no marks]
            table[x={xAx}, y={eigsU}] {geometricalData/eigs.dat};
            
        \addplot+[no marks]
            table[x={xAx}, y={eigsP}] {geometricalData/eigs.dat};
            
        \addplot+[no marks]
            table[x={xAx}, y={eigsE}] {geometricalData/eigs.dat};
        
        \legend{Velocity, Pressure, Energy};
        \end{axis}
            \end{tikzpicture}
    \begin{tikzpicture}
        \begin{axis}[
            name=plot2,
            at=(plot1.right of south east),
            anchor=left of south west,
            xlabel={POD mode no.},
            grid=both,
            grid style={line width=.1pt, draw=gray!10},
            minor tick num=1,
            height=0.49*0.8*\textwidth,
            width=0.44*\textwidth,
            ylabel={Cumulative eigenvalues},
            legend style={at={(0.98,0.3)},anchor=north east},
        ]

        \addplot+[no marks]
            table[x={xAx}, y={eigsNut}] {geometricalData/eigs.dat};
        
        \legend{Eddy viscosity};
        \end{axis}
    \end{tikzpicture}
    \caption{Cumulative eigenvalues trends.}
    \label{fig:eigsGeo}
\end{figure}
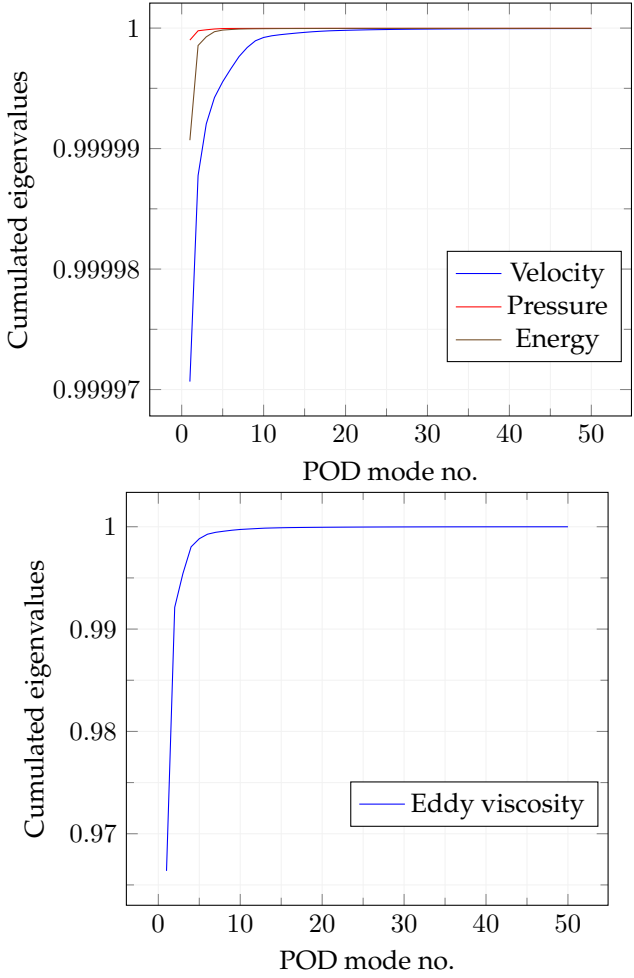
To test the online performances, $20$ new scalar amplitude couples have been randomly selected. $30$ modal basis functions have been picked  for the reconstruction of velocity, pressure and internal energy fields while $15$ modal basis functions have been employed for $\nu_t$. This choice is supported by what is shown in \autoref{fig:eigsGeo}: the increasing trend of the cumulative eigenvalues is pretty fast and this fact allows the discarding of the modes higher than the fixed quantity. For every new parameter couple, the mesh motion has to be performed but the procedure is very efficient since the coefficients for the RBF have to be evaluated and stored just once \cite{aria2021}.
\begin{figure}[!htb]
    \centering
    \begin{tikzpicture}
        \begin{semilogyaxis}[
            xlabel={Epochs},
            ylabel={Loss},
            legend style={at={(0.98,0.98)},anchor=north east},
            grid=both,
            grid style={line width=.1pt, draw=gray!10},
            minor tick num=1,
            height=0.49*0.8*\textwidth,
            width=0.44*\textwidth,
            x tick label style={
                               /pgf/number format/.cd,
                                    fixed,
                               /tikz/.cd
            },
            scaled x ticks=false
        ]
        \addplot+[no marks]
            table[x={xAxL}, y={trainLoss}] {geometricalData/loss3015.dat};
            
        \addplot+[no marks]
            table[x={xAxL}, y={testLoss}] {geometricalData/loss3015.dat};
        
        \legend{Train, Test};
        
        \end{semilogyaxis}
    \end{tikzpicture}
    \caption{Loss function decay for both train and test sets.}
    \label{fig:lossesGeo}
\end{figure}
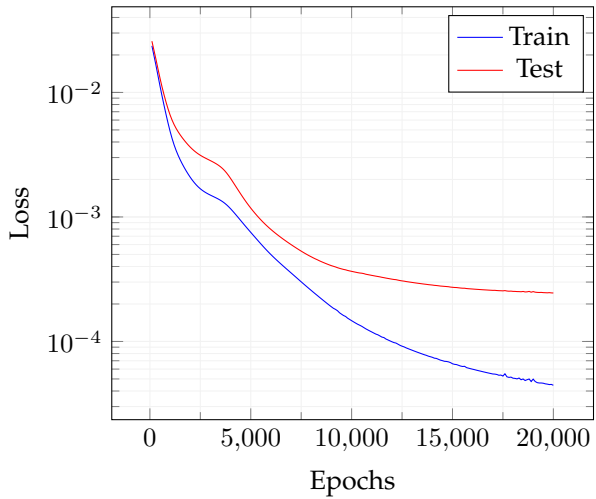
The same neural network used for \autoref{sec:phy} is employed here for what concerns the eddy viscosity. Again, looking at \autoref{fig:lossesGeo}, it can be noticed that the learning of the net seems to stabilize after $2\times 10^4$ epochs which is the threshold we fixed for the training procedure.
\begin{figure}[!htb]
    \centering
    \begin{tikzpicture}
        \begin{axis}[
            name=plot1,
            xlabel={Parameter no.},
            ylabel={$L^2$ relative error},
            legend style={at={(0.98,0.5)},anchor=north east},
            grid=both,
            grid style={line width=.1pt, draw=gray!10},
            minor tick num=1,
            height=0.49*0.8*\textwidth,
            width=0.44*\textwidth,
        ]

        \addplot+[only marks]
            table[x={parameter}, y={errorU}] {geometricalData/errors.dat};
        \addplot+[only marks]
            table[x={parameter}, y={errorP}] {geometricalData/errors.dat};
        \addplot+[only marks]
            table[x={parameter}, y={errorE}] {geometricalData/errors.dat};
        
        \legend{Velocity, Pressure, Energy};
        \end{axis}
            \end{tikzpicture}
    \begin{tikzpicture}
        \begin{axis}[
            name=plot2,
            at=(plot1.right of south east),
            anchor=left of south west,
            xlabel={Parameter no.},
            ylabel={$L^2$ relative error},
            legend style={at={(0.98,0.8)},anchor=north east},
            grid=both,
            grid style={line width=.1pt, draw=gray!10},
            minor tick num=1,
            height=0.49*0.8*\textwidth,
            width=0.44*\textwidth,
        ]

        \addplot+[only marks]
            table[x={parameter}, y={errorNut}] {geometricalData/errors.dat};
        
        \legend{Eddy viscosity};
        \end{axis}
    \end{tikzpicture}
    \caption{$L^2$ norm relative errors.}
    \label{fig:errorsGeo}
\end{figure}
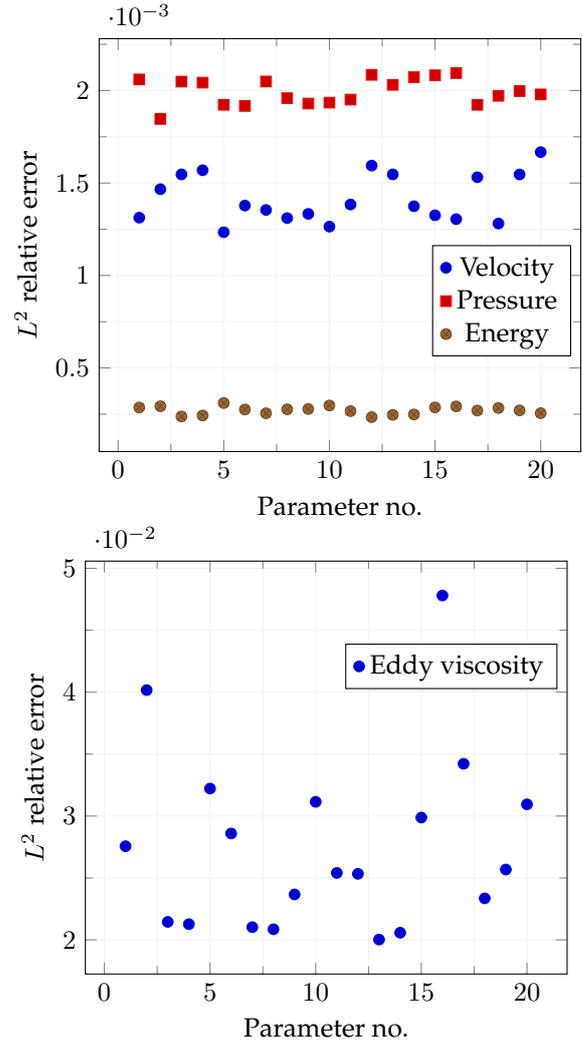
The resulting $L^2$ norm errors for all the parameter couples in the online set are shown in \autoref{fig:errorsGeo}. Once again a discrepancy of about one order of magnitude can be noticed between the relative errors for the quantities of interest and the one calculated for the eddy viscosity. This is because we are using a very simple and small network but it reveals to be reliable enough to make the online algorithm work fine.
\begin{figure}[!htb]
    \centering
    \begin{tabular}{cc}
    \includegraphics[width=0.32 \textwidth]{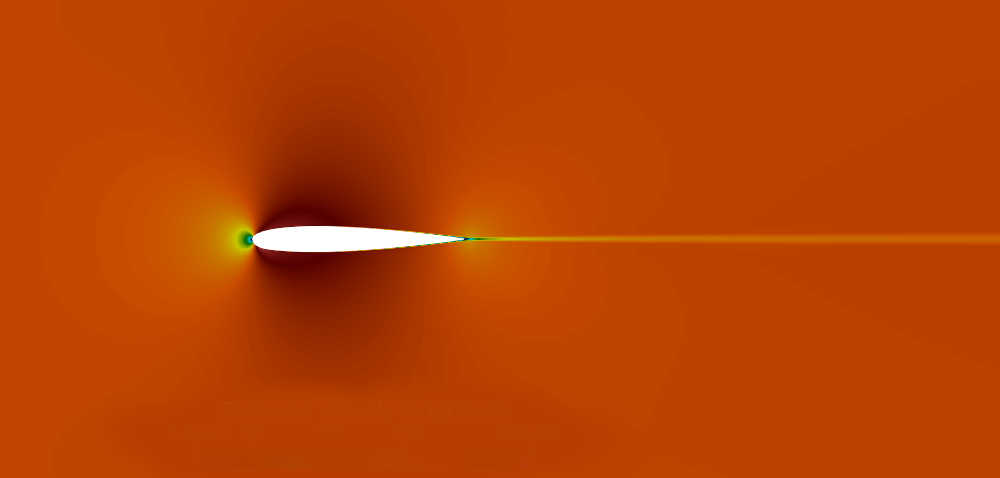}
    \includegraphics[width=0.32 \textwidth]{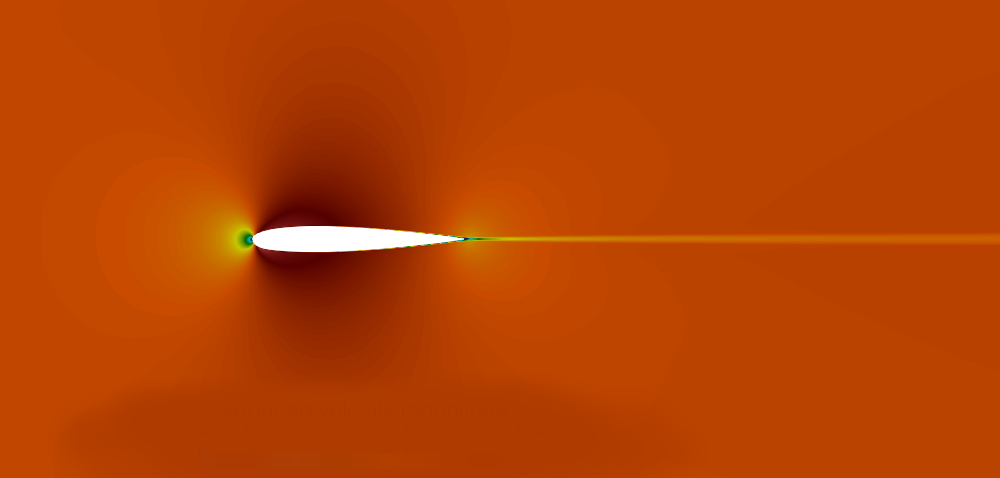}\\
    \includegraphics[width=0.32 \textwidth]{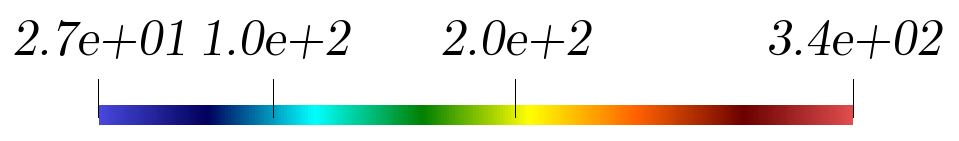}
    \end{tabular}
    \begin{tabular}{cc}
    \includegraphics[width=0.32 \textwidth]{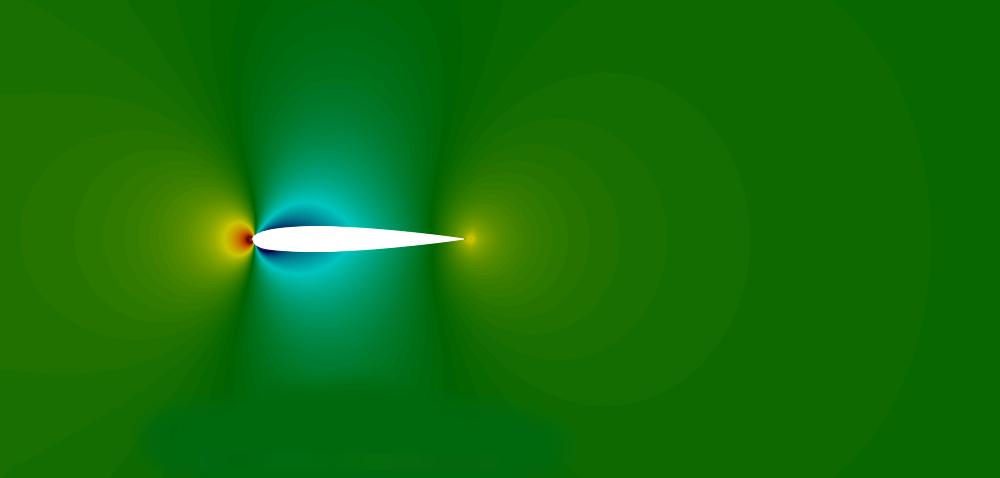}
    \includegraphics[width=0.32 \textwidth]{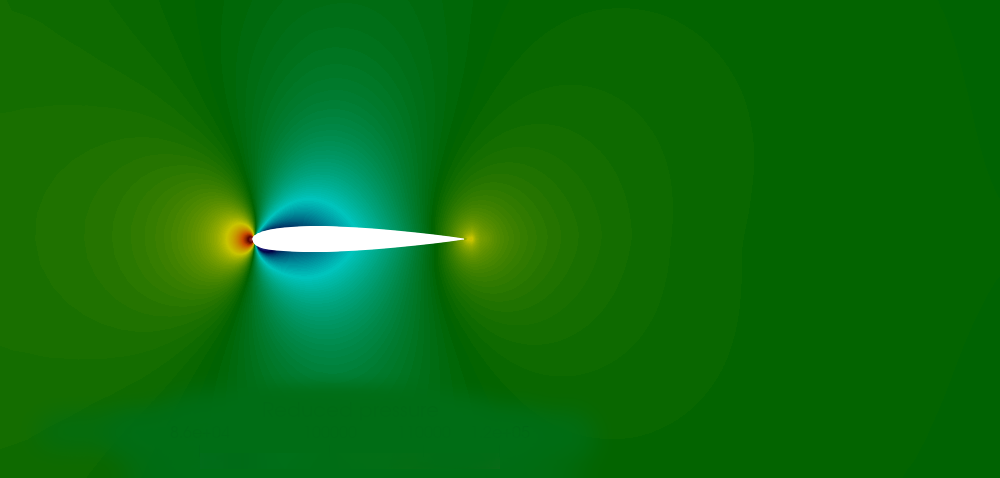}\\
    \includegraphics[width=0.32 \textwidth]{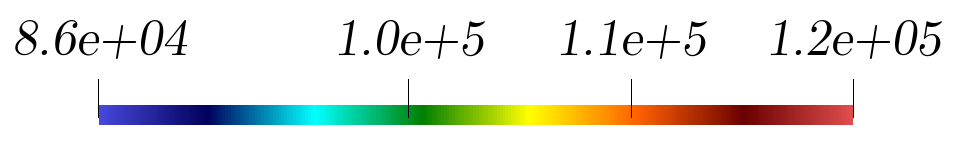}
    \end{tabular}
    \begin{tabular}{cc}
    \includegraphics[width=0.32 \textwidth]{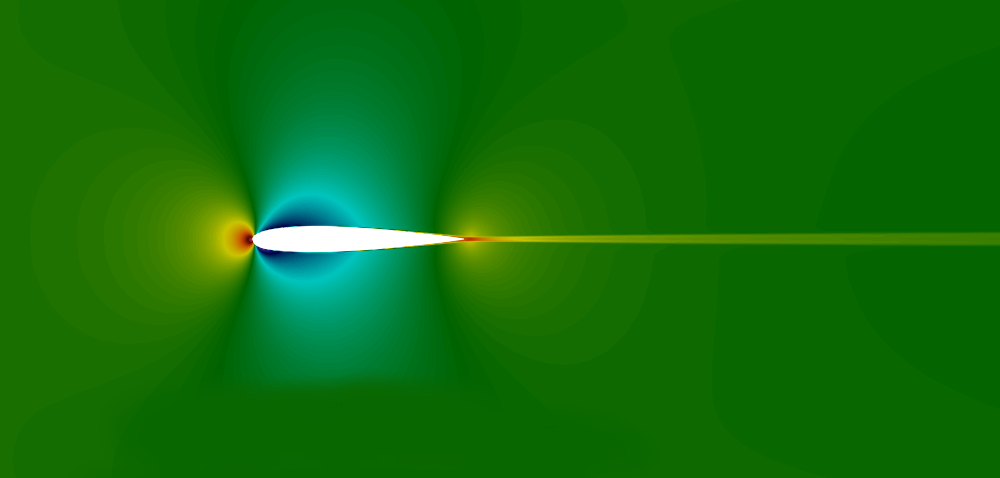}
    \includegraphics[width=0.32 \textwidth]{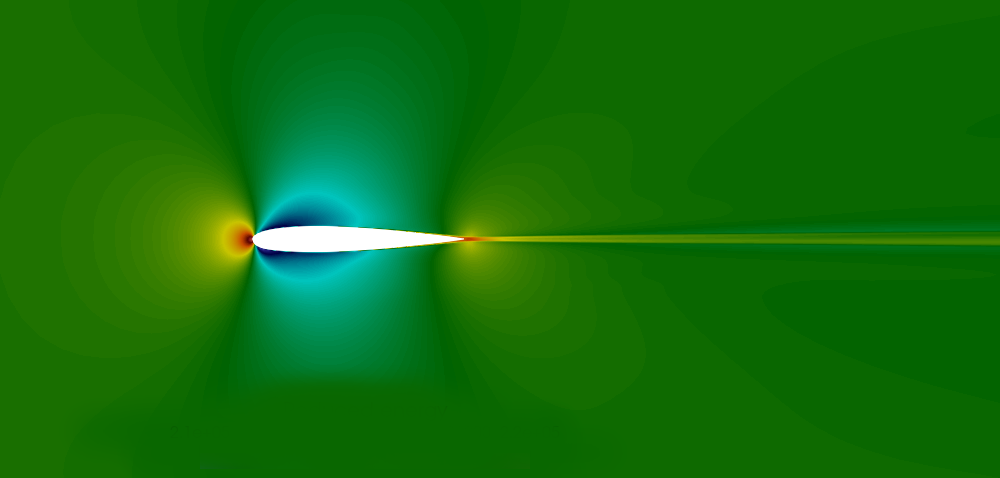} \\   
    \includegraphics[width=0.32 \textwidth]{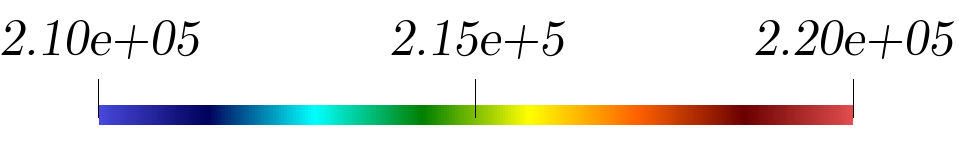}
    \end{tabular}
    \caption{Comparison between full-order (first column) and reduced-order (second column) solutions: \textit{velocity} magnitude (first row), \textbf{pressure} (second row) and \textit{energy} (third row). These fields refer to the resolution of the problem for $\pi_{top}\simeq0.004$ and $\pi_{bottom}\simeq0.086$ which has been selected as a random value in the online parameter set.}
    \label{fig:geometricalFields1}
\end{figure}
\begin{figure}[H]
    \centering
    \includegraphics[width=0.32 \textwidth]{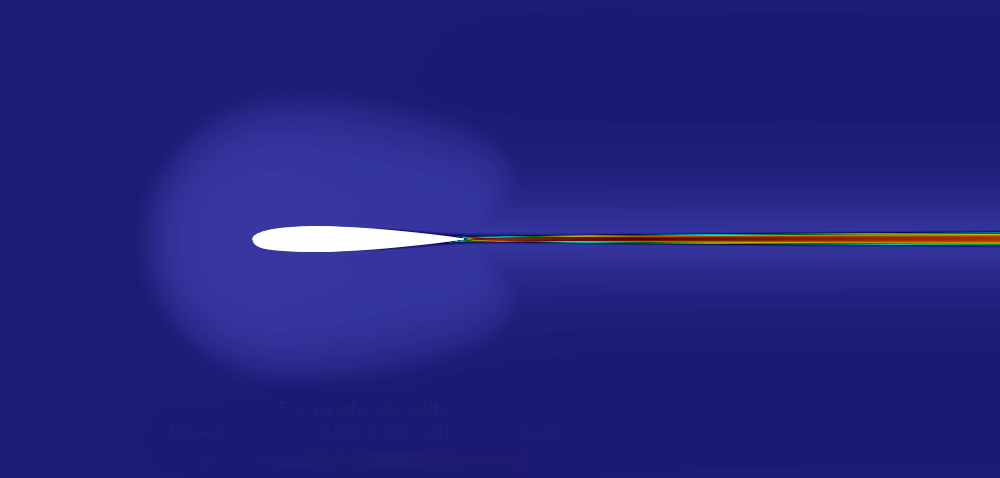}
    \includegraphics[width=0.32 \textwidth]{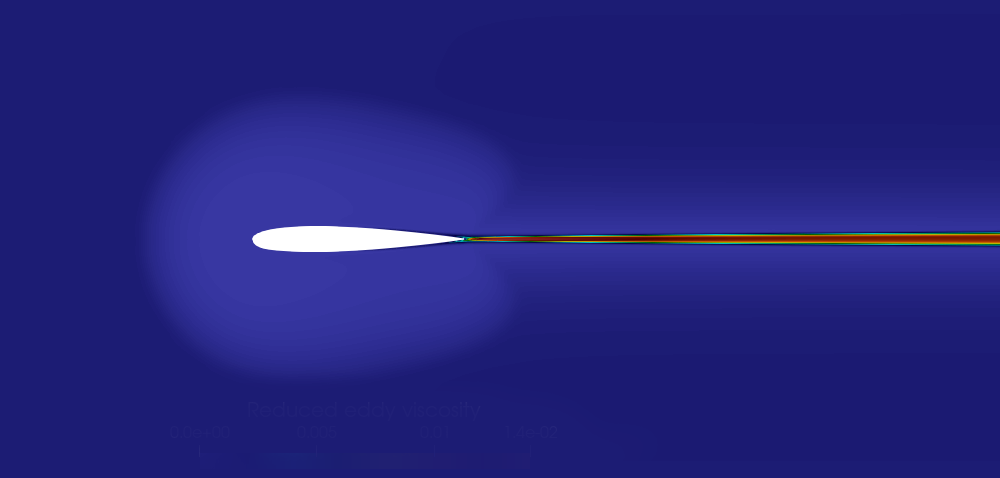}\\
    \includegraphics[width=0.32 \textwidth]{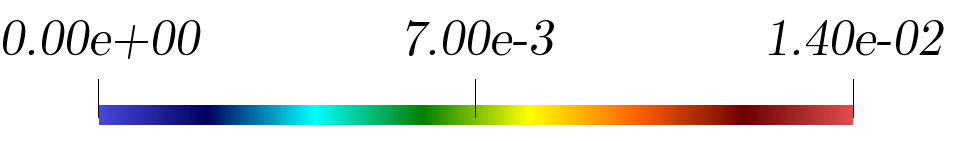}
    \caption{Comparison between full-order (left picture) and reduced-order (right picture) for the eddy viscosity solutions. These fields refer to the resolution of the problem for $\pi_{top}\simeq0.004$ and $\pi_{bottom}\simeq0.086$ which has been selected as a random value in the online parameter set.}
    \label{fig:geometricalNut1}
\end{figure}
\begin{figure}[H]
    \centering
    \begin{tabular}{cc}
    \includegraphics[width=0.32 \textwidth]{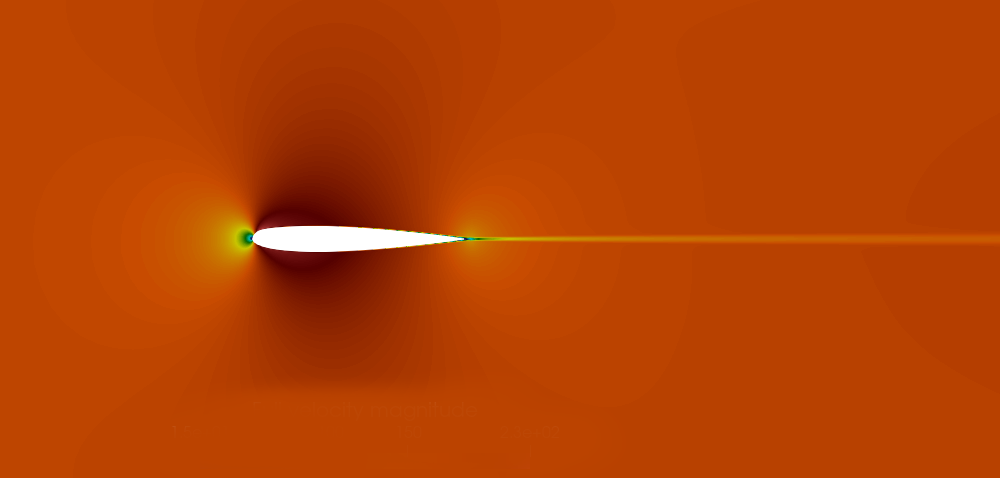}
    \includegraphics[width=0.32 \textwidth]{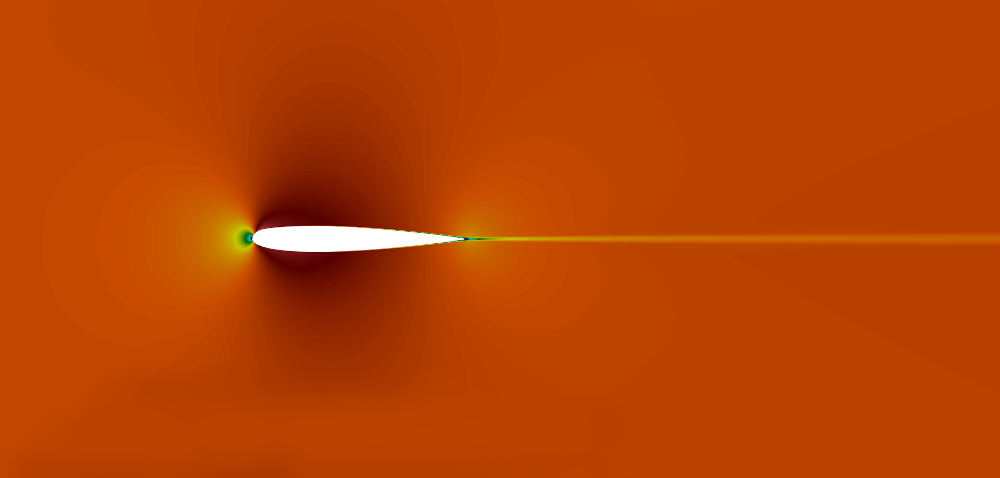}\\
    \includegraphics[width=0.32 \textwidth]{geometricalData/img/0Uscale_crop.png}
    \end{tabular}
     \begin{tabular}{cc}
    \includegraphics[width=0.32 \textwidth]{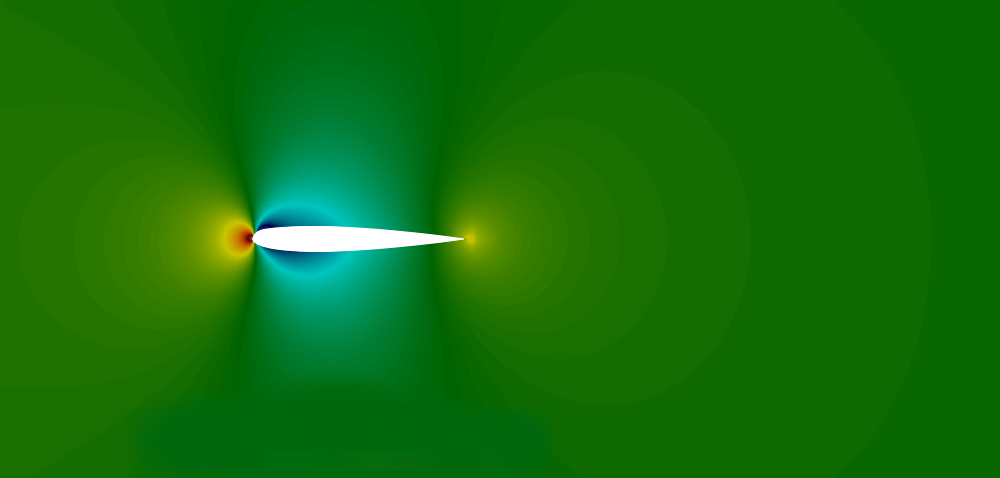}
    \includegraphics[width=0.32 \textwidth]{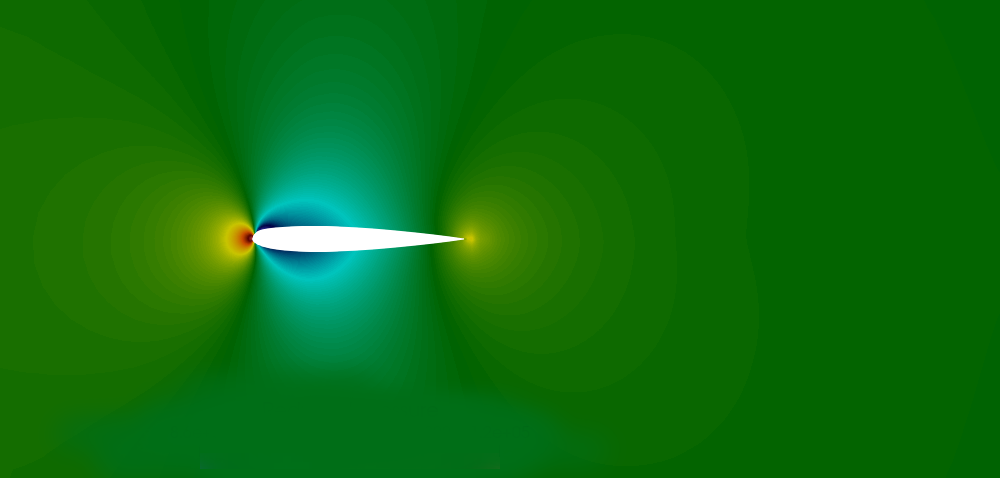}\\
    \includegraphics[width=0.32 \textwidth]{geometricalData/img/0Pscale_crop.png}
    \end{tabular}
     \begin{tabular}{cc}
    \includegraphics[width=0.32 \textwidth]{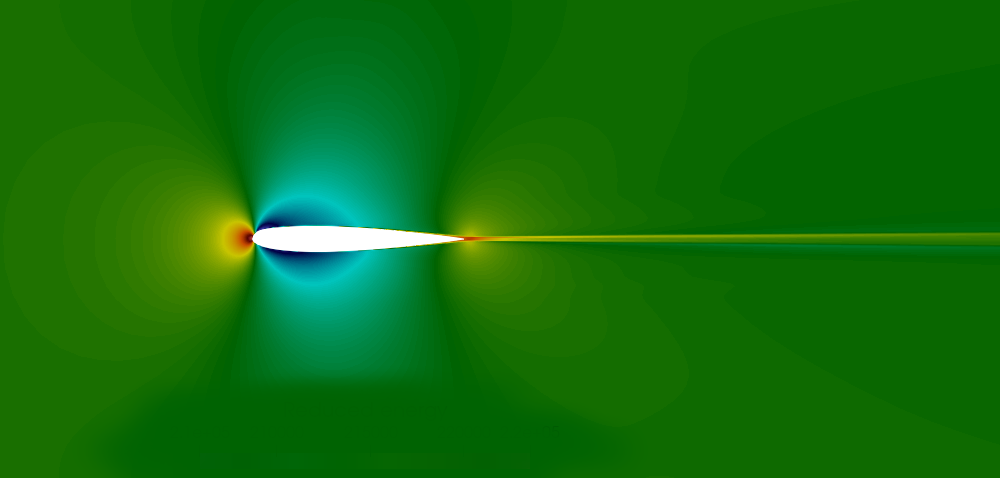}
    \includegraphics[width=0.32 \textwidth]{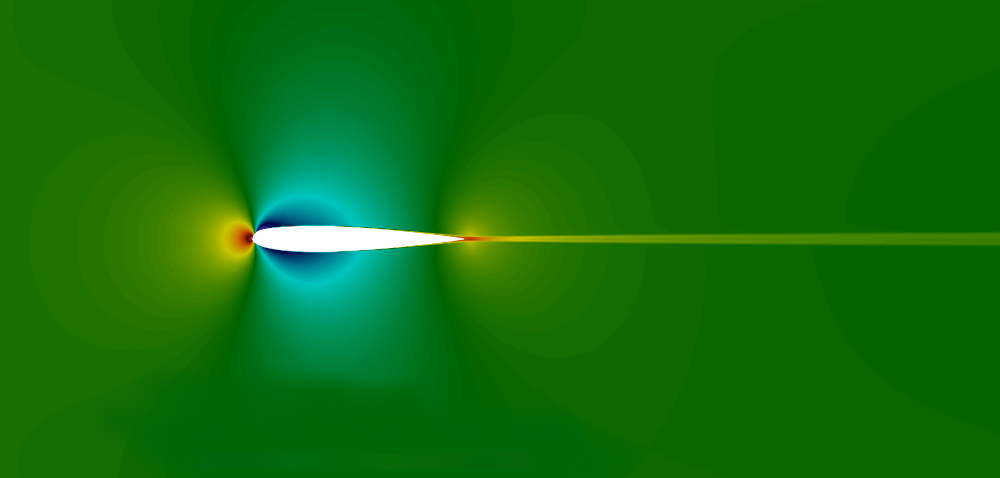} \\   
    \includegraphics[width=0.32 \textwidth]{geometricalData/img/0Escale_crop.png}
    \end{tabular}
    \caption{Comparison between full-order (first column) and reduced-order (second column) solutions: \textit{velocity} magnitude (first row), \textit{pressure} (second row) and \textit{energy} (third row). These fields refer to the resolution of the problem for $\pi_{top}\simeq 0.095$ and $\pi_{bottom}\simeq 0.003$ which has been selected as a random value in the online parameter set.}
    \label{fig:geometricalFields2}
\end{figure}
\begin{figure}[!htb]
    \centering
    \includegraphics[width=0.32 \textwidth]{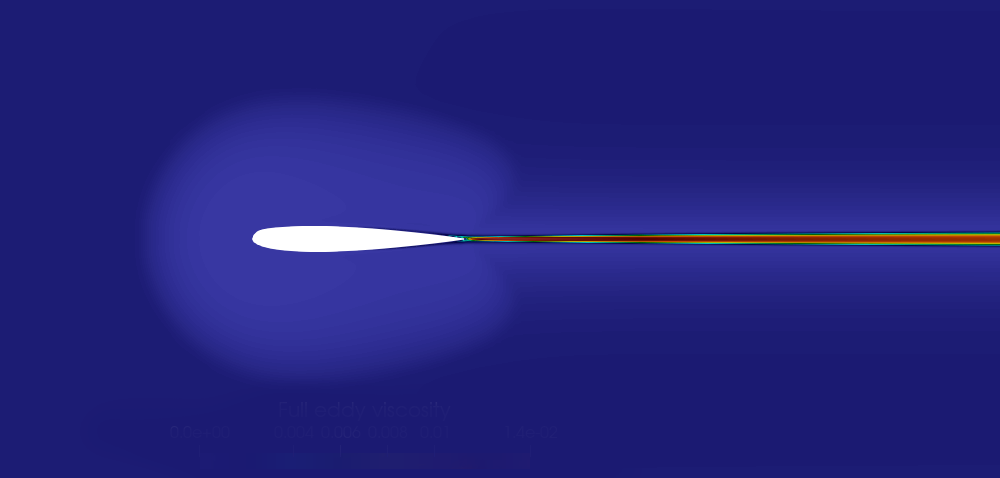}
    \includegraphics[width=0.32 \textwidth]{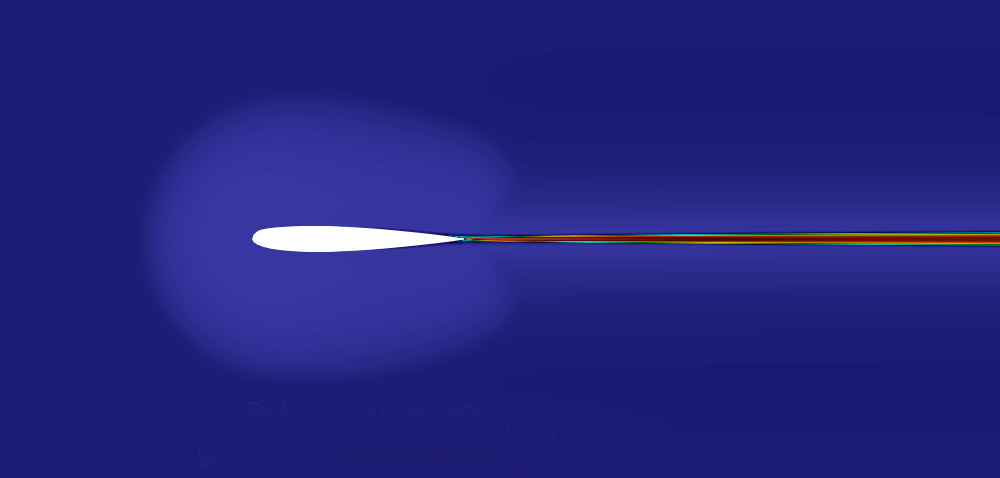}\\
    \includegraphics[width=0.32 \textwidth]{geometricalData/img/0nutscale_crop.png}
    \caption{Comparison between full-order (left picture) and reduced-order (right picture) for the  eddy viscosity solutions. These fields refer to the resolution of the problem for $\pi_{top}\simeq0.095$ and $\pi_{bottom}\simeq0.003$ which has been selected as a random value in the online parameter set.}
    \label{fig:geometricalNut2}
\end{figure}
In \autoref{fig:geometricalFields1}, \autoref{fig:geometricalNut1}, \autoref{fig:geometricalFields2} and \autoref{fig:geometricalNut2} a comparison between offline and online solutions is depicted for two different parameter couples selected from the online set. Even if the two solutions are obtained for airfoil geometries that are perturbed in opposite directions, in both cases the method exhibits good reliability properties even though the intermediate solutions introduce in the snapshots matrix could be highly inaccurate and trigger instabilities in the ROMs. 

%% file: sections/conclusions.tex
\section{Conclusions and future perspectives}\label{sec:conclusions}
This study focused on compressible flows by proposing a new mixed technique, capable of merging the reliability of Galerkin-projection methods together with the versatility of data-driven strategies in turbulence and compressible flows. The good results obtained for both a physical and geometrical parameterized benchmarks make this approach quite promising. From one hand,  the possibility to freely select the turbulence model avoiding the necessity of changing the whole architecture is attractive, while on the other hand, the guarantee of a strong connection with physical aspects given by the projection of conservation laws is reassuring.

The segregated compressible algorithm proposed in \autoref{sec:redComp},  also introduces a way to provide accurate reduced solutions without any kind of stabilization: the employment of a decoupled approach for the compressible turbulent Navier-Stokes equations relies on the chipping of the saddle point formulation. For this reason no stabilization for pressure is required: as shown in both \autoref{sec:phy} and \autoref{sec:geom}, pressure field solutions do not exhibit significant instability or inaccuracy issues. This aspect helps the procedure on being more consistent without pollution of the resulting solution due to stabilization.
A natural extension of this work will be a deep  analysis with others existing approaches both in the methodology and application. Another extension will be the application of neural networks to  approximate the functional evaluations required by the online phase to overtake the necessity of reconstructing the full fields at each iteration. This aspect would increase the performances but it has to be carefully calibrated to avoid possible drifting of the algorithm resulting on the loss of the convergence.

A final aspect that can be improved is the neural network itself: a weighted strategy where eigenvalues play a relevant role in the loss function would, in principle, enhance the training stage since the first modal basis functions, represented by the highest eigenvalues, are the  most significant ones on the reconstruction procedure. 

%% file: sections/aknw.tex
\section*{Acknowledgments}
This research has been supported by the European Union Funding for Research and Innovation -Horizon 2020 Program- in the framework of European Research Council Executive Agency: Consolidator Grant H2020 ERC CoG 2015 AROMA-CFD project 681447 "Advanced Reduced Order Methods with Applications in Computational Fluid Dynamics" (PI Prof. Gianluigi Rozza) and by the H2020 MSCA RISE ARIA (grant 872442) project. We also acknowledge the INDAM-GNCS project "Tecniche Numeriche Avanzate per Applicazioni Industriali", the support by MIUR (Italian Ministry for University and Research) FARE-X-AROMA-CFD project and PRIN "Numerical Analysis for Full and Reduced-Order Methods for Partial Differential Equations" (NA-FROM-PDEs). The main computations in this work were carried out by the usage of ITHACA-FV \cite{ithacasite}, an implementation in OpenFOAM \cite{ofsite} for reduced-order modelling techniques. We acknowledge developers and contributors of each of the aforementioned libraries.
%